\documentclass[journal,12pt,onecolumn,draftclsnofoot]{IEEEtran}
\usepackage{graphicx}
\usepackage{amsmath}
\DeclareMathOperator*{\argmax}{arg\,max}
\DeclareMathOperator*{\argmin}{arg\,min}
\usepackage{textcomp}
\usepackage{amssymb}
\usepackage{relsize}
\usepackage{eso-pic}
\usepackage{caption}
\usepackage[font=footnotesize,caption=false,subrefformat=parens,labelformat=parens]{subfig}
\usepackage{array,multirow}
\usepackage{float}
\usepackage{color}
\usepackage{url}
\usepackage{graphicx}
\usepackage{amsfonts}
\usepackage{lipsum}  
\usepackage[normalem]{ulem}
\usepackage{color}
\usepackage{cite}
\usepackage{epstopdf}
\usepackage{bm}
\usepackage{mathtools}
\usepackage[algoruled,resetcount,linesnumbered]{algorithm2e}
\usepackage{array}
\usepackage{multirow}
\usepackage{multicol}
\usepackage{mathtools}
\usepackage{tablefootnote}
\newcolumntype{P}[1]{>{\centering\arraybackslash}p{#1}}

\usepackage{lipsum}
\usepackage{fancyhdr} 
\graphicspath{ {fig/} } 

\begin{document}

\title{RF-Based Low-SNR Classification of UAVs Using Convolutional Neural Networks}
\author{\normalsize Ender Ozturk, Fatih Erden, and Ismail Guvenc
\thanks{This work has been supported in part by NASA under the Federal Award ID number NNX17AJ94A.}%
\thanks{Ender Ozturk (eozturk2@ncsu.edu), Fatih Erden (ferden@ncsu.edu), and Ismail Guvenc (iguvenc@ncsu.edu) are with the Department of Electrical and Computer Engineering, North Carolina State University, Raleigh, NC 27606.}
\vspace{-6mm}
}
\maketitle
\begin{abstract}
This paper investigates the problem of classification of unmanned aerial vehicles~(UAVs) from radio frequency~(RF) fingerprints at the low signal-to-noise ratio~(SNR) regime. We use convolutional neural networks~(CNNs) trained with both RF time-series images and the spectrograms of 15 different off-the-shelf drone controller RF signals. When using time-series signal images, the CNN extracts features from the signal transient and envelope. As the SNR decreases, this approach fails dramatically because the information in the transient is lost in the noise, and the envelope is distorted heavily. In contrast to time-series representation of the RF signals, with spectrograms, it is possible to focus only on the desired frequency interval, i.e., 2.4~GHz ISM band, and filter out any other signal component outside of this band. These advantages provide a notable performance improvement over the time-series signals-based methods. To further increase the classification accuracy of the spectrogram-based CNN, we denoise the spectrogram images by truncating them to a limited spectral density interval. Creating a single model using spectrogram images of noisy signals and tuning the CNN model parameters, we achieve a classification accuracy varying from 92\% to 100\% for an SNR range from $-$10~dB to 30~dB, which significantly outperforms the existing approaches to our best knowledge.
\end{abstract}


\section{Introduction}
Unmanned aerial vehicles~(UAVs) or drones have recently gained a great deal of interest among researchers due to unrivaled commercial opportunities in various fields, such as wireless communications, logistics, delivery, search and rescue, smart agriculture, surveillance, among others~\cite{shakhatreh2019unmanned}. In addition, the recent COVID-19 outbreak revealed the importance of remote operations in every aspect of life, which may accelerate social acceptance of drone use cases such as delivery of goods and medication~\cite{riananda2020smart,lin2018drone,chamola2020comprehensive}. With the new advances in airspace regulations and drone-related technologies, it is expected that there will be more and more UAVs in the skies for various use cases, sharing the airspace with other aerial vehicles. 

Innate advantages of UAVs that make them popular, such as ease of operation and low cost, could also be considered as major disadvantages from a security perspective. There have been many criminal activities recently with drones involved, 
and their small sizes make it difficult to detect, classify, and interdict them~\cite{hard_to_detect,guvenc2018detection}. In this regard, Federal Aviation Agency (FAA) of the United States recently announced a Proposed Rule that elaborates the future action that would require remote identification of unmanned aircraft systems to address safety and security concerns~\cite{FAA_proposed_rule}.

\begin{figure}[!t]
\centering\vspace{-3mm}
\centerline{\includegraphics[width=0.60\columnwidth]{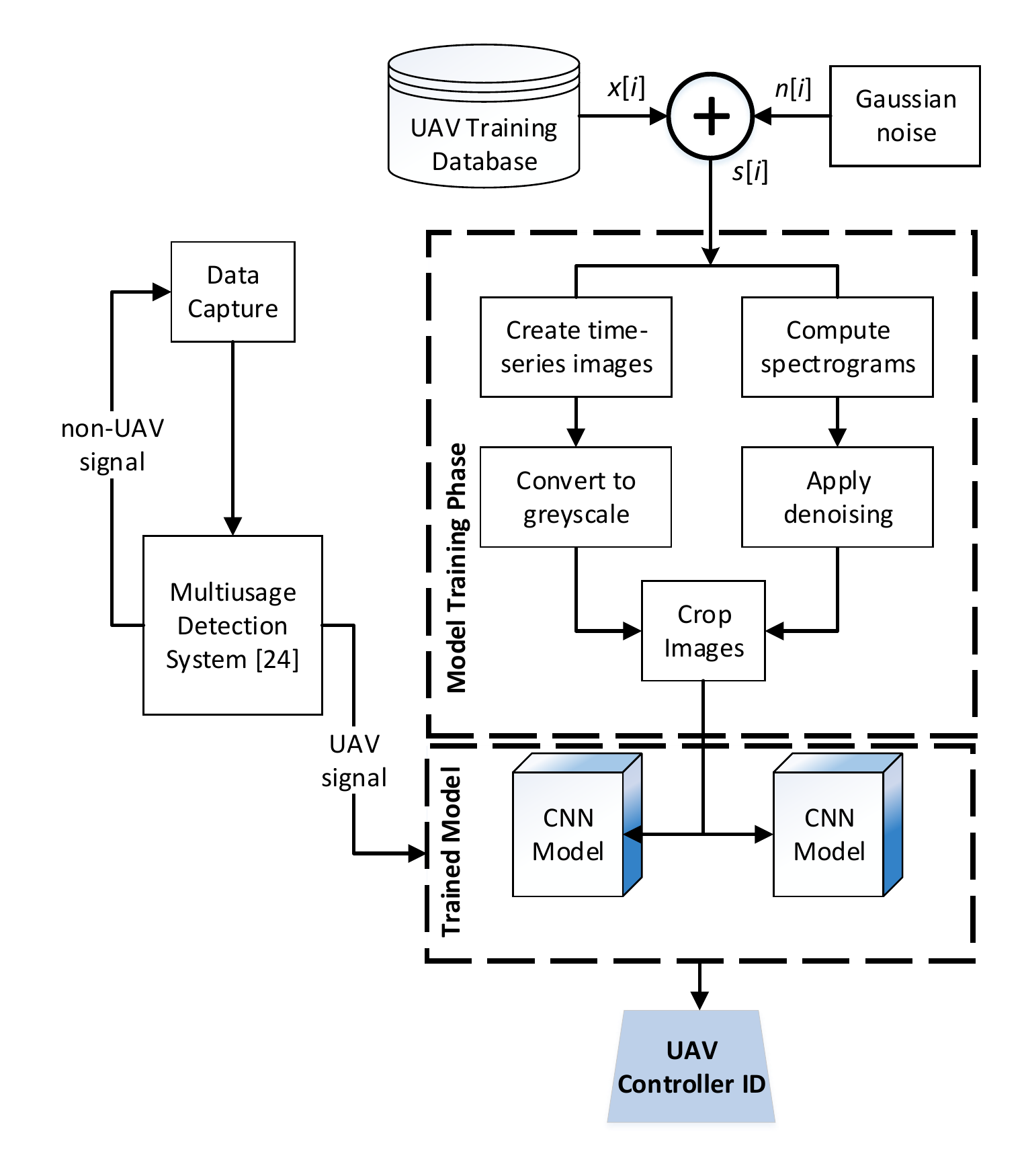}}
\caption{Overview of the proposed system. Multistage detector classifies the captured data as of type \emph{UAV} or \emph{non-UAV}. In the case of a \emph{UAV} signal, captured data is artificially noised, and time-series and spectrogram images are created afterwards for training the corresponding CNN models. Time-series images are converted to grayscale to increase computational efficiency. Spectrograms are denoised to increase the model accuracy. Separate CNN models are trained and predictions are made using these CNN models.
} 

\label{Fig:flowchart}
\vspace{-4mm}
\end{figure}
\begin{table*}[t!]
\renewcommand\arraystretch{1.3}
\caption{Related work on detection and classification of drones using ML techniques.}
\begin{tabular}{P{1.2cm}P{2cm}P{2.5cm}P{3cm}P{1.5cm}P{1cm}P{2.1cm}P{1.3cm}}
\hline
\textbf{Literature} & \textbf{Source type}& \textbf{Features} &\textbf{Data process method}& \textbf{Classification} &\textbf{\# of UAVs} &  \textbf{Accuracy} & \textbf{Noise consideration}\\
\hline
\hline
\cite{ANN_drone_comm_signal}& Drone RF signals & Slope, kurtosis, skewness & Several ML algorithms&\text{\sffamily X}&N/A& 96.36\%&\text{\sffamily X}\\
\hline
\cite{channel_state}& Drone RF signal& CSI data & Channel state information &\text{\sffamily X}&N/A& 86.6\%&\text{\sffamily X}\\
\hline
\cite{aucustic_SVM} & Acoustic waves & MFCC and LPCC & SVM &\text{\sffamily X}&N/A& 96.7\%&\text{\sffamily X}\\
\hline
\cite{acoustic_spectrogram}& Acoustic waves & STFT features & CNN &\text{\sffamily X}&N/A& 99.87\%&\text{\sffamily X}\\
\hline
\cite{photo_image_CNN_kNN}& Camera images & RGB arrays & CNN for moving body detection and kNN for detection&\text{\sffamily X}&N/A& 93\%&\text{\sffamily X}\\
\hline
\cite{a_study_cnn_photo_image}& Camera images & RGB arrays & CNN on ZF and VGG16 and Fast R-CNN &\text{\sffamily X}&N/A& 0.66 mAP&\text{\sffamily X}\\
\hline
\cite{ren_and_jiang} & Radar signals & Spectrogram & 2-D complex-log-Fourier transform & \text{\sffamily X} & N/A &3.27\% EER &\text{\sffamily X}\\
\hline
\cite{messina_snr_as_feature} & Radar signals & Range Doppler Matrix & SVM & \text{\sffamily X} & N/A &98\% &\text{\sffamily X}\\
\hline
\cite{micro_doppler_dualband_Radar} & Radar signals & Micro-Doppler signature & PCA feature extraction on spectrograms & \checkmark & 3 &94.7\% &\text{\sffamily X}\\
\hline
\cite{Huizing_2019_miniUAV_magazine} & Radar signals & Micro-Doppler spectrogram & CNN and LSTM-RNN &\checkmark&5& 97.7\% &\text{\sffamily X}\\
\hline
\cite{Kim_2017_merged_doppler}&Radar signals & Micro-Doppler signature & CNN &\checkmark&6& 94.7\% &\text{\sffamily X}\\
\hline
\cite{micro-doppler_EMD}& Radar signals & Micro-Doppler signatures through EMD & SVM &\checkmark&11& $>$95\%&\text{\sffamily X}\\
\hline
\cite{Molchanov}&Radar signals & Micro-Doppler signatures & SVM & \checkmark & 11 & 95.4\% & \text{\sffamily X}\\
\hline
\cite{SNR_2}&Radar signals&Range Doppler spectrum&CNN&\text{\sffamily X}&N/A&99.5\% and 54.2\% for 0~dB&\checkmark\\
\hline
\cite{IEEE_TIFS_8}&Drone RF signals&Statistical features e.g., mean, median, RMS&Logistic regression&\checkmark&8&88-94\% in 0.35s&\text{\sffamily X} \\
\hline
\cite{SNR_1}&Radar signals&Micro-Doppler signature&ANN on MLP&\checkmark&4&Various&\checkmark\\
\hline
\cite{mpact_rffingerprints} & Controller RF signals & Shape factor, kurtosis, variance  & Several ML algorithms&\checkmark&17& 98.13\% and 40\% for 0~dB SNR&\checkmark\\
\hline
\textbf{This work} & \textbf{Controller RF signals} & \textbf{Time-series signal and spectrogram RGB arrays} & \textbf{CNN} &\textbf{\checkmark}&\textbf{15}& \textbf{99.7}\% \textbf{and 99.5}\% \textbf{for 0~dB} \textbf{SNR}&\textbf{\checkmark}\\
\cline{1-8}
\end{tabular}\label{tab:literature_comparison}
\vspace{-2mm}
\end{table*}
UAVs can be identified through a set of features that uniquely represent them. These features can be extracted from various data sources, such as visual data, acoustic, RF, or radar signals.
Each of these source types has its own pros and cons which we will review in the next section. In this study, we develop a convolutional neural network (CNN)-based classifier using both time-series signal images and spectrogram images of 15 different drone controller RF signals to classify drones of different makes and models. We use controller signals as the dataset was already in possession; however, proposed approach can also be directly applied to the signals transmitted from drones to their controllers.
Flowchart of the overall procedure is given in Fig.~\ref{Fig:flowchart}. 

For the classification tasks that involve RF fingerprinting, variations in the signal-to-noise ratio~(SNR) of the received RF signals is a challenging problem. In this work, we also address this practical problem by considering a range of SNR levels from $-$10~dB to 30~dB while training the CNN models. Noisy training data is generated by adding artificial white noise to the original data. When using spectrogram images to train the CNN models, we only focus on the frequency range of interest, which improves classification accuracy significantly in comparison with time-series images. We also apply denoising on the spectrogram images to further improve the performance at low SNRs. We tune the spectral density level that will appear on the spectrogram image and filter out spectral densities lower than the tuned level. Our proposed classifier highly outperforms previously published work, especially at low SNRs. \looseness = -1

A possible use case of such a system would be about the upcoming FAA regulation on Remote ID~\cite{FAA_proposed_rule}. Remote ID is defined as the ability of a UAV to provide the relevant identity information to other parties. Even the drones will be obliged to reveal their IDs to comply with this regulation, it will still be possible for the malicious drones to fake their IDs. The system proposed in this work can be a part of a framework that verifies the drone IDs and make sure that the flying drone has the same ID as in the FAA's logs. This way counter measures can be taken in the presence of a threat. 

The rest of the paper is organized as follows. In Section II, a comprehensive literature review including the information of noise consideration is given. In Section III, the dataset and the procedure for obtaining noisy samples are introduced. Section IV discusses image data preprocessing step and the CNN-based classifier used in this work. Experimental results and relevant discussions are presented in Section V. Finally, the paper is concluded in Section VI.

\section{Literature Review}
Various approaches have been proposed in the literature for detection and classification of drones. In Table~\ref{tab:literature_comparison}, we summarize the related literature on drone detection and classification with some representative work and emphasis on the number of UAVs considered, classification accuracy, and noise considerations. Here we use the term \textit{detection} as a special case of classification that has only two classes (i.e., UAV/non-UAV). Techniques used to achieve these tasks can be categorized based on the type of the data being captured (e.g., radar signals, drone or controller radio frequency~(RF) signals, acoustic data, or camera images), features extracted from the data (e.g., RF fingerprints, spectrogram images), and the machine learning~(ML) algorithms deployed for classification. Acoustic sensors do not require line-of-sight (LOS); however, they suffer from short range, as drones could operate very quietly~\cite{aucustic_SVM, acoustic_mar2020}, and data gathered using microphone systems are prone to wind and environmental clutter. On the other hand, a LOS vision under daylight is essential for techniques that utilize camera images~\cite{photo_image_CNN_kNN, visual_turkish}. Using thermal or laser-based cameras to overcome this issue increases the cost significantly. 

\begin{figure*}[t]
\centering
\captionsetup[subfigure]{labelformat=parens}
\subfloat[]{\includegraphics[width=0.191\linewidth]{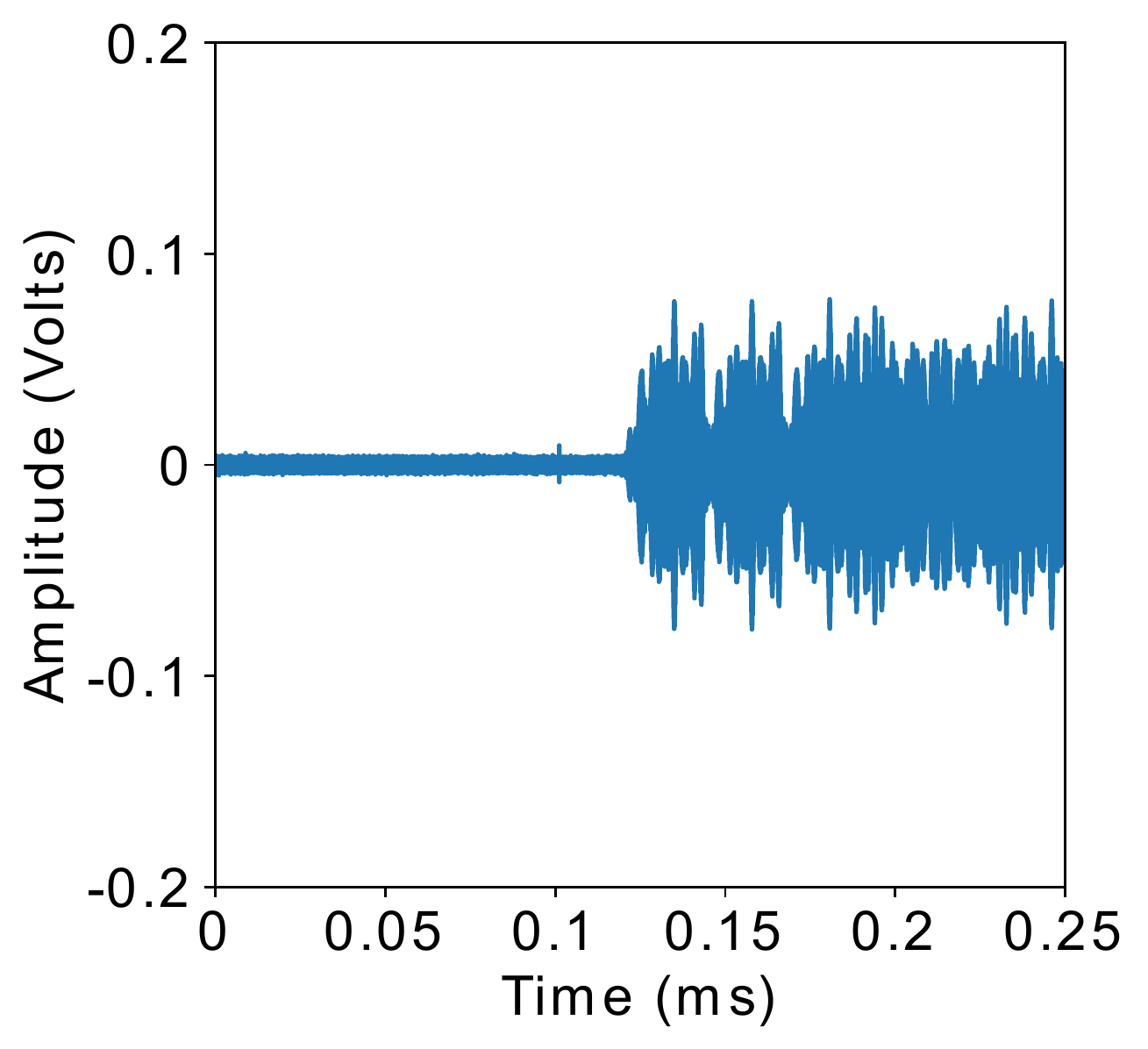}}
\subfloat[]{\includegraphics[width=0.191\linewidth]{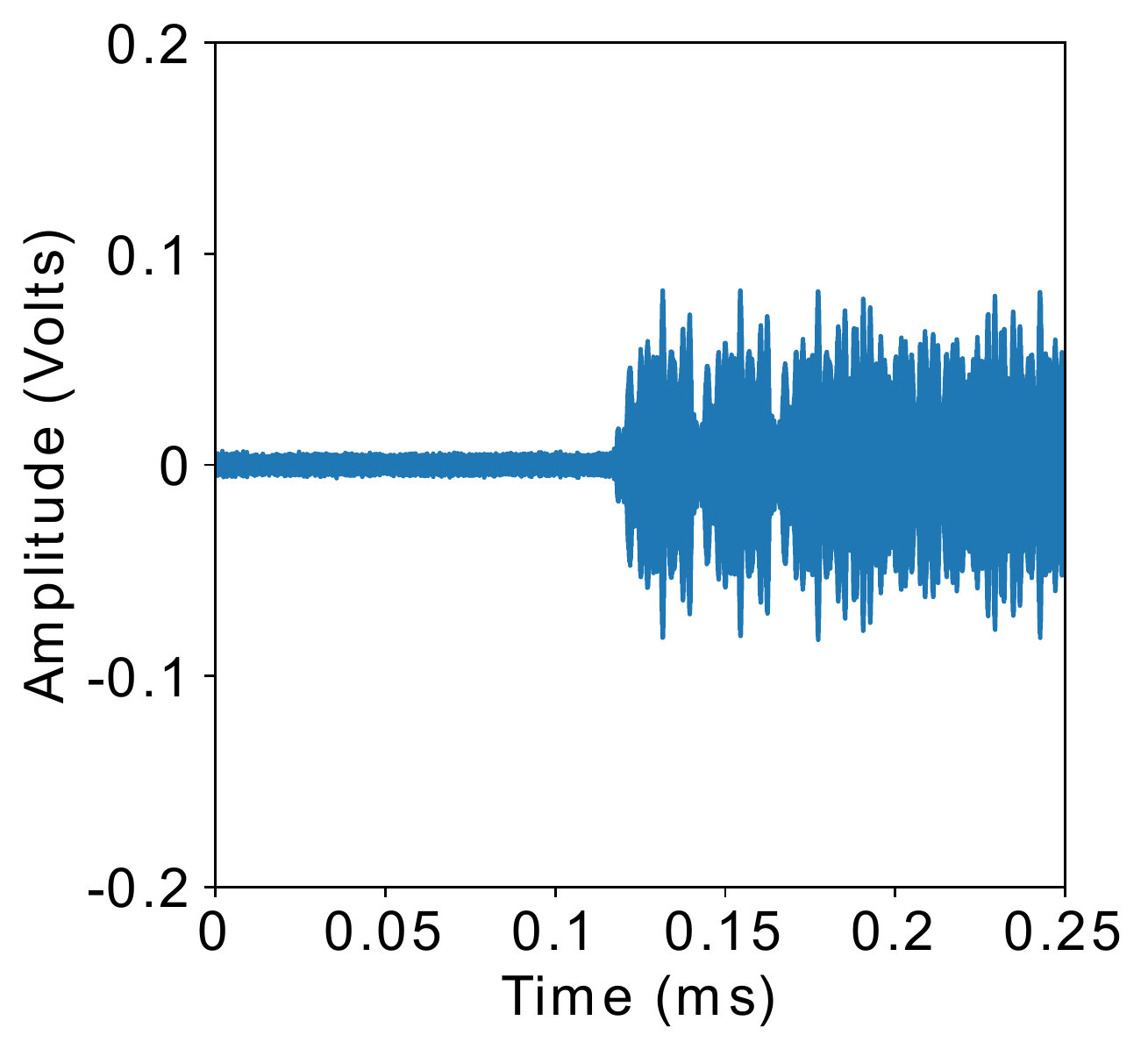}}
\subfloat[]{\includegraphics[width=0.191\linewidth]{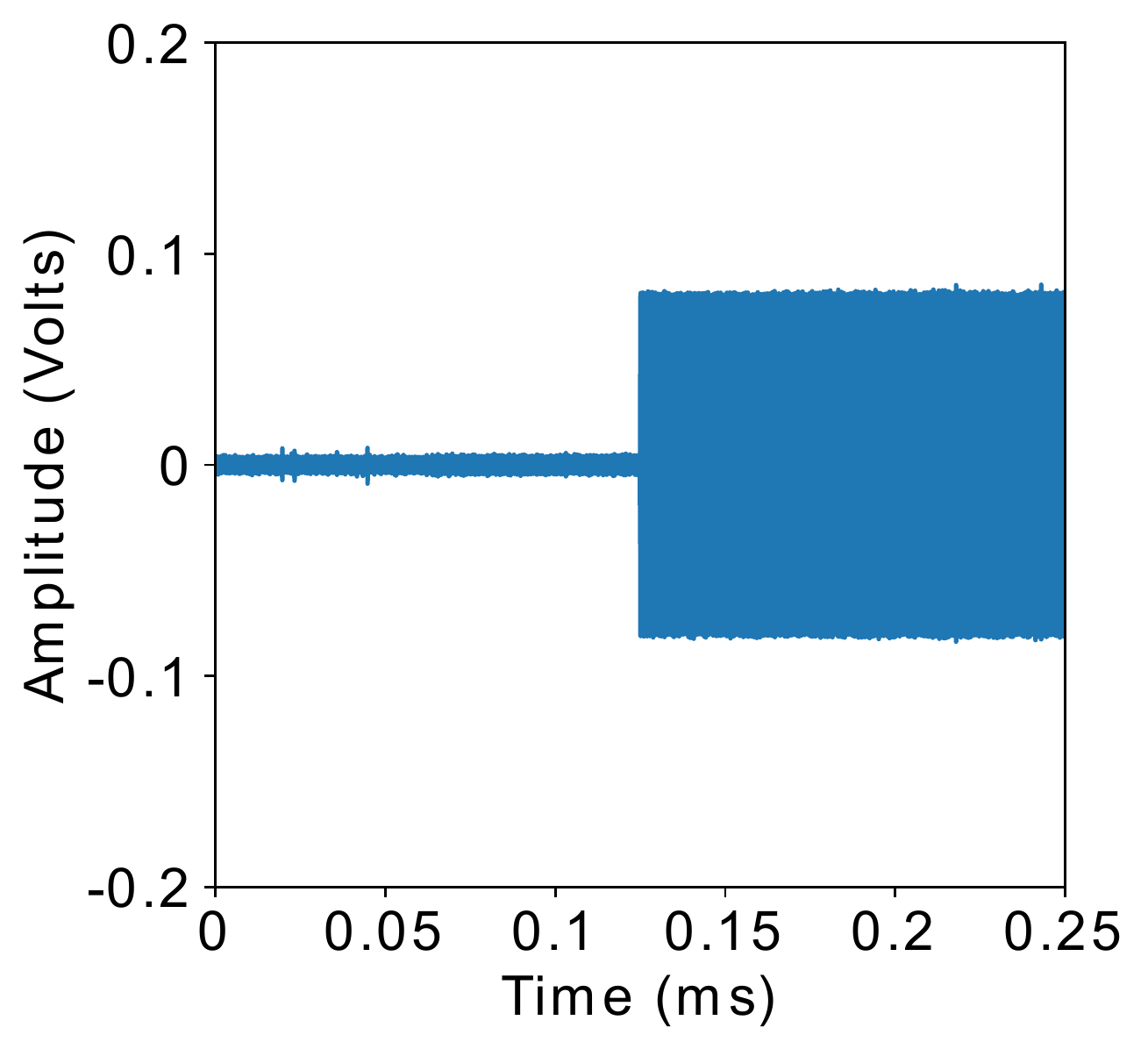}}
\subfloat[]{\includegraphics[width=0.191\linewidth]{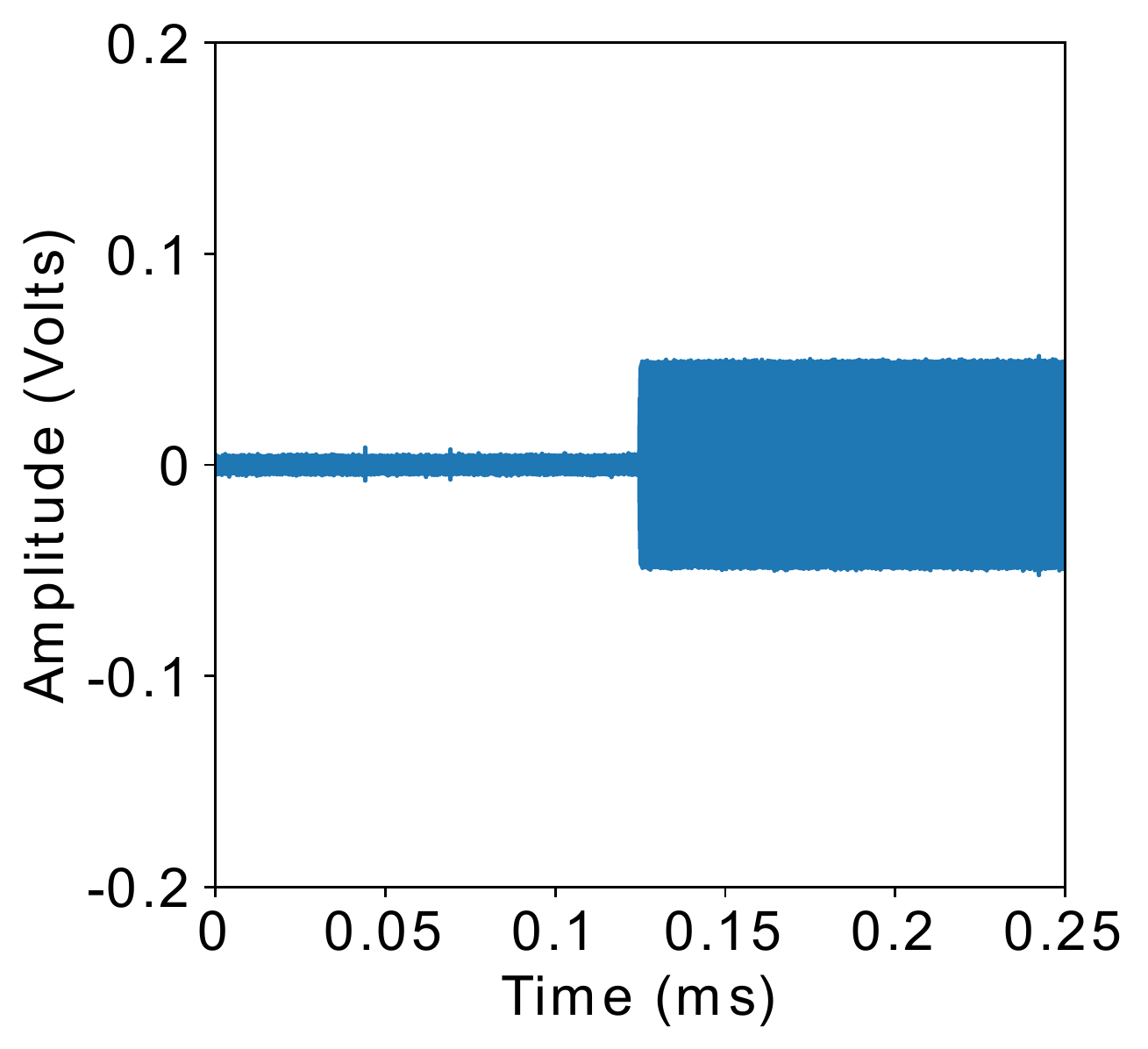}}
\subfloat[]{\includegraphics[width=0.191\linewidth]{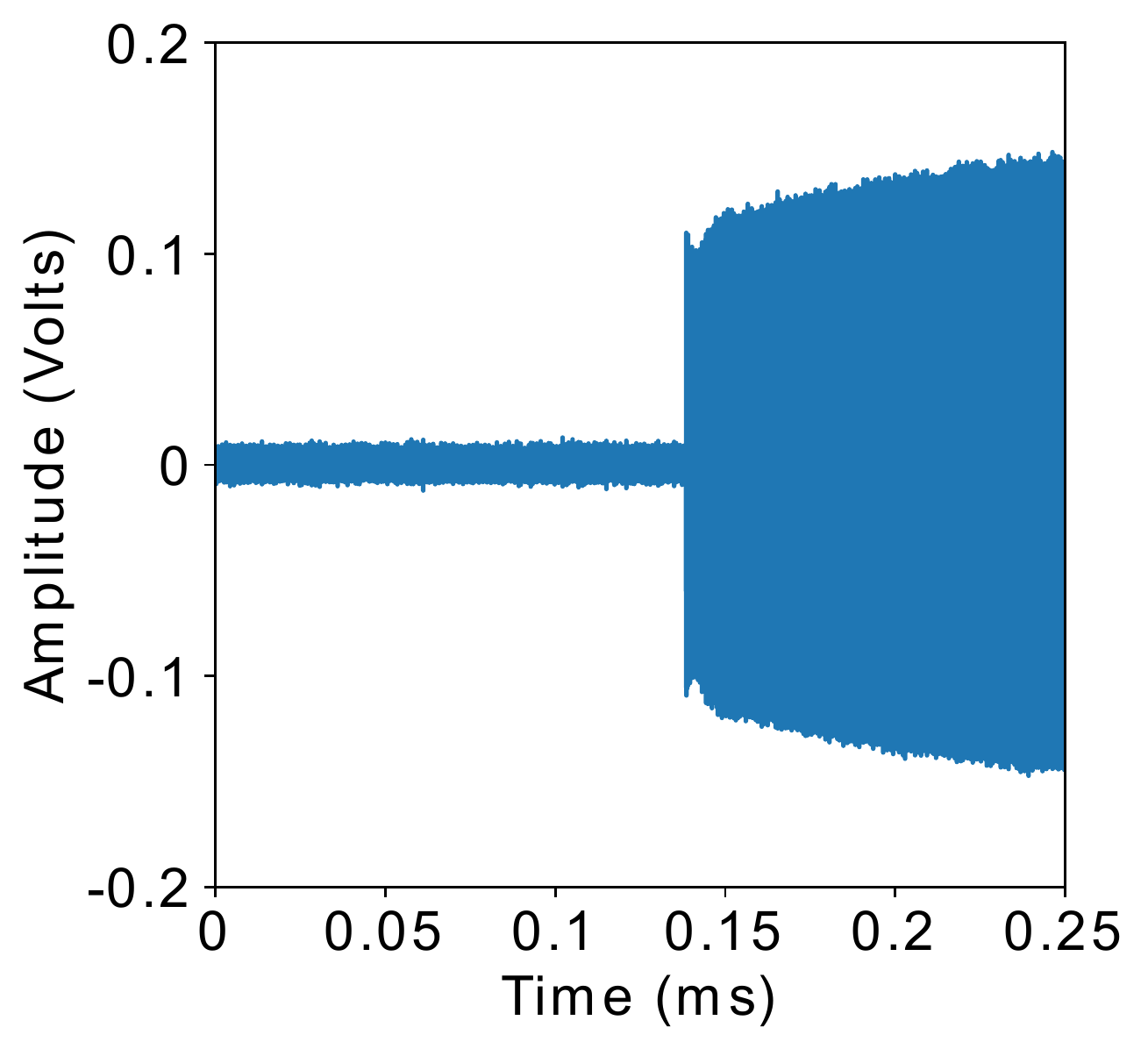}}\newline
\caption{Sample controller time-series RF signals: (a) DJI Matrice 100, (b) DJI Matrice 600, (c) Spektrum DX5e, (d) FlySky FS-T6, and (e) Spektrum JR X9303.
RF signals from different controllers may look alike, making it difficult to identify the drones based on only the envelopes of the captured signals. 
}
\label{Fig:sample_plots}
\end{figure*}

Radar signals are immune to environmental factors, such as acoustic noise and fog. However, drones are small devices with tiny propellers which make it hard to perceive and distinguish them from each other by most radars. A high-frequency wideband radar could be used to deal with these difficulties~\cite{Kim_2017_merged_doppler, micro_doppler_extraction, Choi_microdoppler_short, micro_doppler_dualband_Radar}. Such radars are considerably expensive and suffer from high path loss. RF signals of either drones themselves or controllers are mostly at \mbox{sub-6}~GHz band and share unlicensed Wi-Fi bands. As a result of this, equipment to capture RF signals are affordable, but on the downside, RF-based techniques require special attention for handling interference from other co-channel signal sources. Besides, no LOS is required, and these techniques are immune to many problems that acoustic and visual techniques suffer from.\looseness=-1

RF signals can be used for classification of the UAVs, either directly or indirectly after some processing. In~\cite{mpact_rffingerprints,ANN_drone_comm_signal, IEEE_TIFS_8}, time-domain statistical properties of the RF signal, such as slope, kurtosis, skewness, shape factor and variance, are used as features along with different ML algorithms to detect and classify drones. However, since unlicensed bands are heavily employed, time-domain information suffers from low SNR. Frequency-domain representation of RF signals can also be used to distinguish between different types of drones. Transforming RF signals into the frequency domain filters out the out-of-band noise and helps improve classification accuracy up to a certain extent.   
In the literature, 
there are studies using radar signals and spectrograms to detect and classify drones~\cite{ren_and_jiang,micro_doppler_dualband_Radar,Huizing_2019_miniUAV_magazine,Kim_2017_merged_doppler,micro-doppler_EMD,Molchanov,SNR_2}. However,  there is no study that utilize spectrograms of RF signals in the context of UAV detection/classification to the best of our knowledge.

Even though the mass majority of classification efforts in this field aim to identify drone make and model to support a decision of friend/foe, there are some other work that use ML techniques to identify drone pilots. For example, in~\cite{drone_pilot_classificaton}, drone controller RF signals are recorded to characterize pilot activity, and different types of maneuvers that a pilot could do are used as features. 

Classification accuracy should be considered together with the number of UAVs as it gets harder to classify UAVs with high accuracy as the number of classes increases. For studies which have \text{\sffamily X} marks in the \textit{Classification} column, the proposed models performed only \textit{detection} which means there are only two classes. We also provide the information about whether the work considers noise or not, to better emphasize our contribution. \looseness=-1 


\section{Dataset and Noising Procedure}\label{sec:problem}
In this work, the dataset in~\cite{mpact_rffingerprints} is used. This dataset consists of RF signals from 15 different off-the-shelf UAV controllers listed in Table~\ref{tab:controllers_list}. RF signals were captured using an ultra-wideband~(UWB) antenna and an oscilloscope with a sampling rate of 20~Gsa/s. Total number of samples in each signal is $5\times10^6$, which corresponds to a time duration of 250~\textmu s. Time-series and spectrogram images are created from the training RF signals, and CNN models are generated for each image database.
\subsection{Image Creation Process}
\label{sec:Images}
\begin{figure*}[t]
\centering
\captionsetup[subfigure]{labelformat=empty}
\subfloat[]{\includegraphics[width=0.3\linewidth]{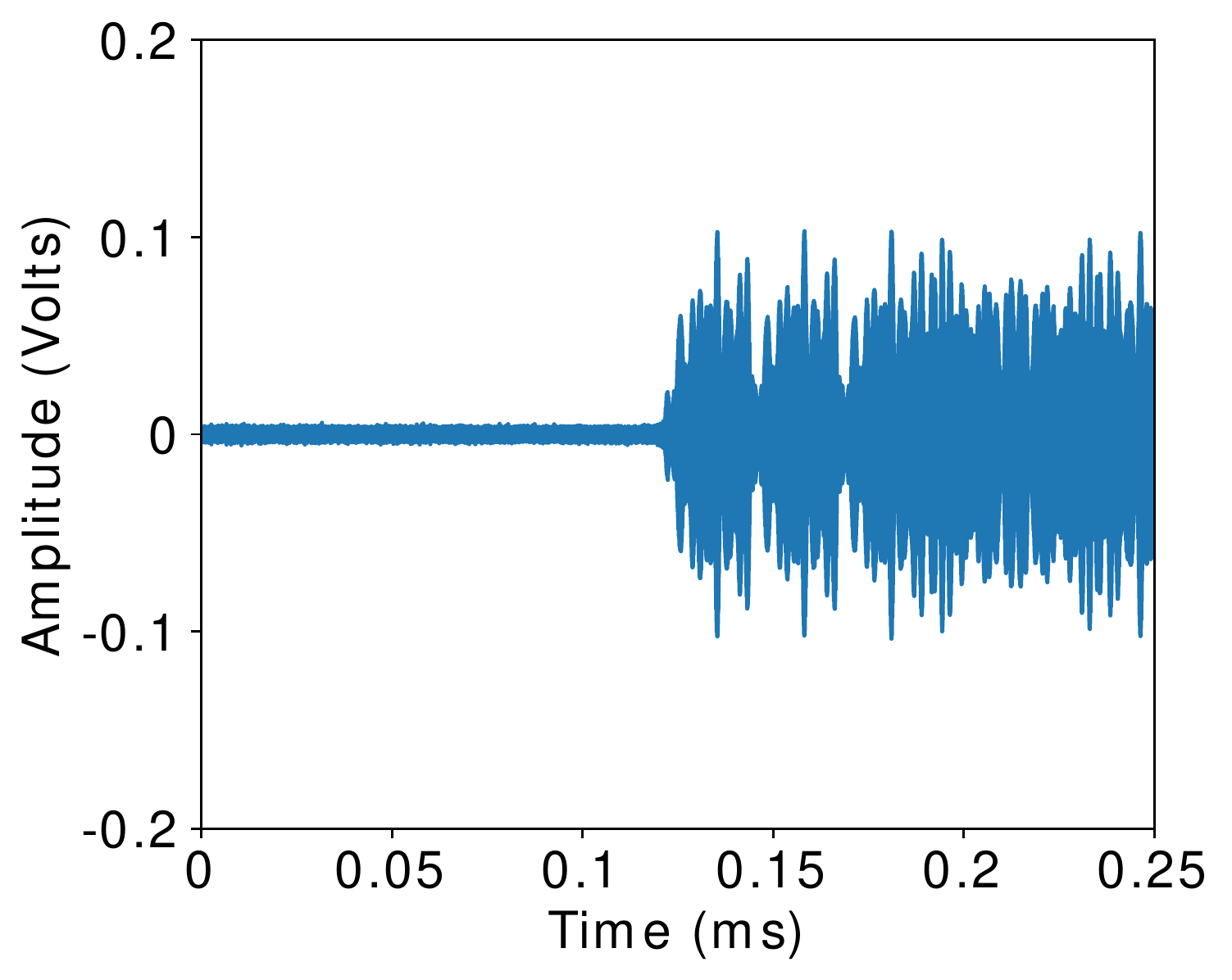}}
\subfloat[]{\includegraphics[width=0.3\linewidth]{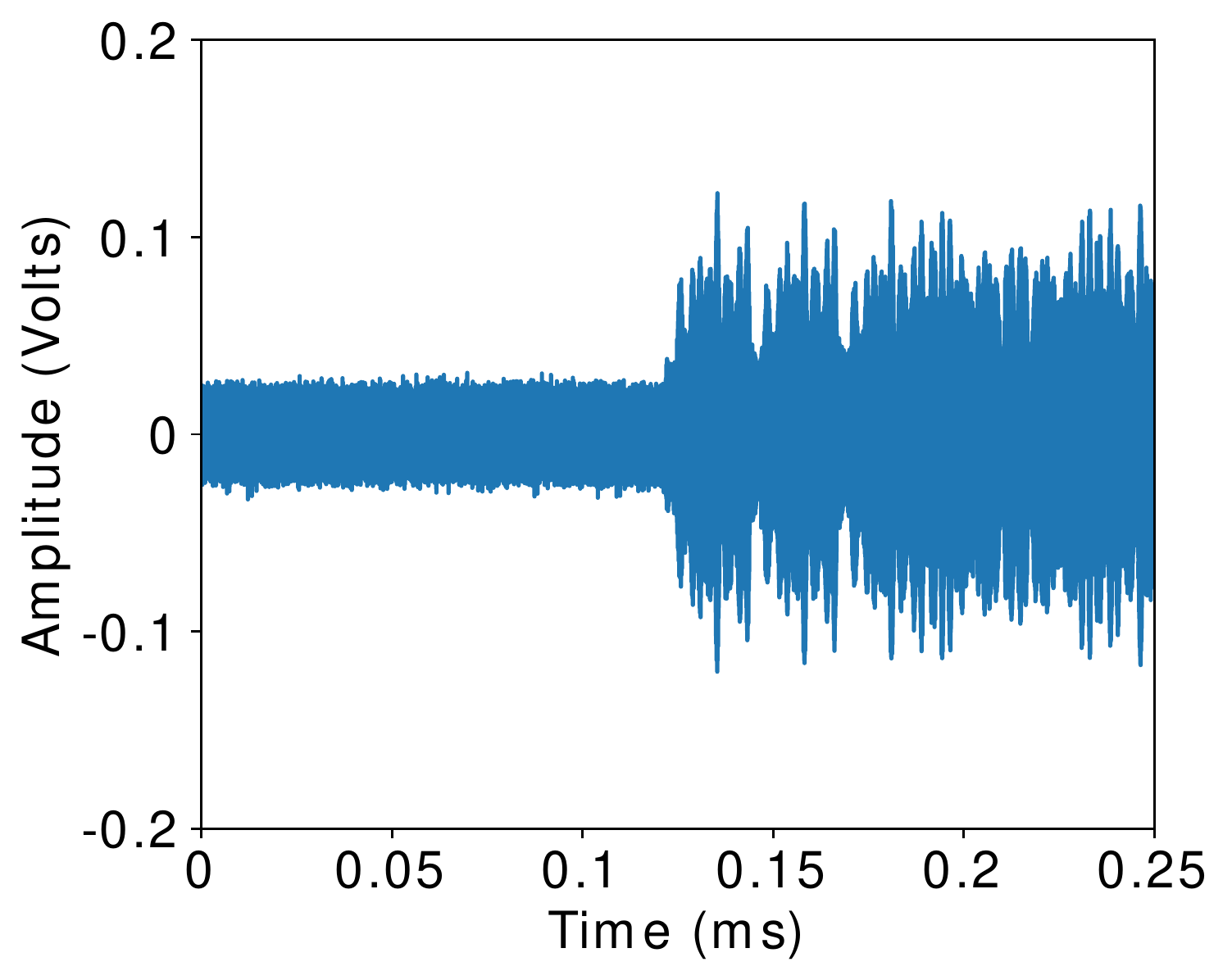}}
\subfloat[]{\includegraphics[width=0.3\linewidth]{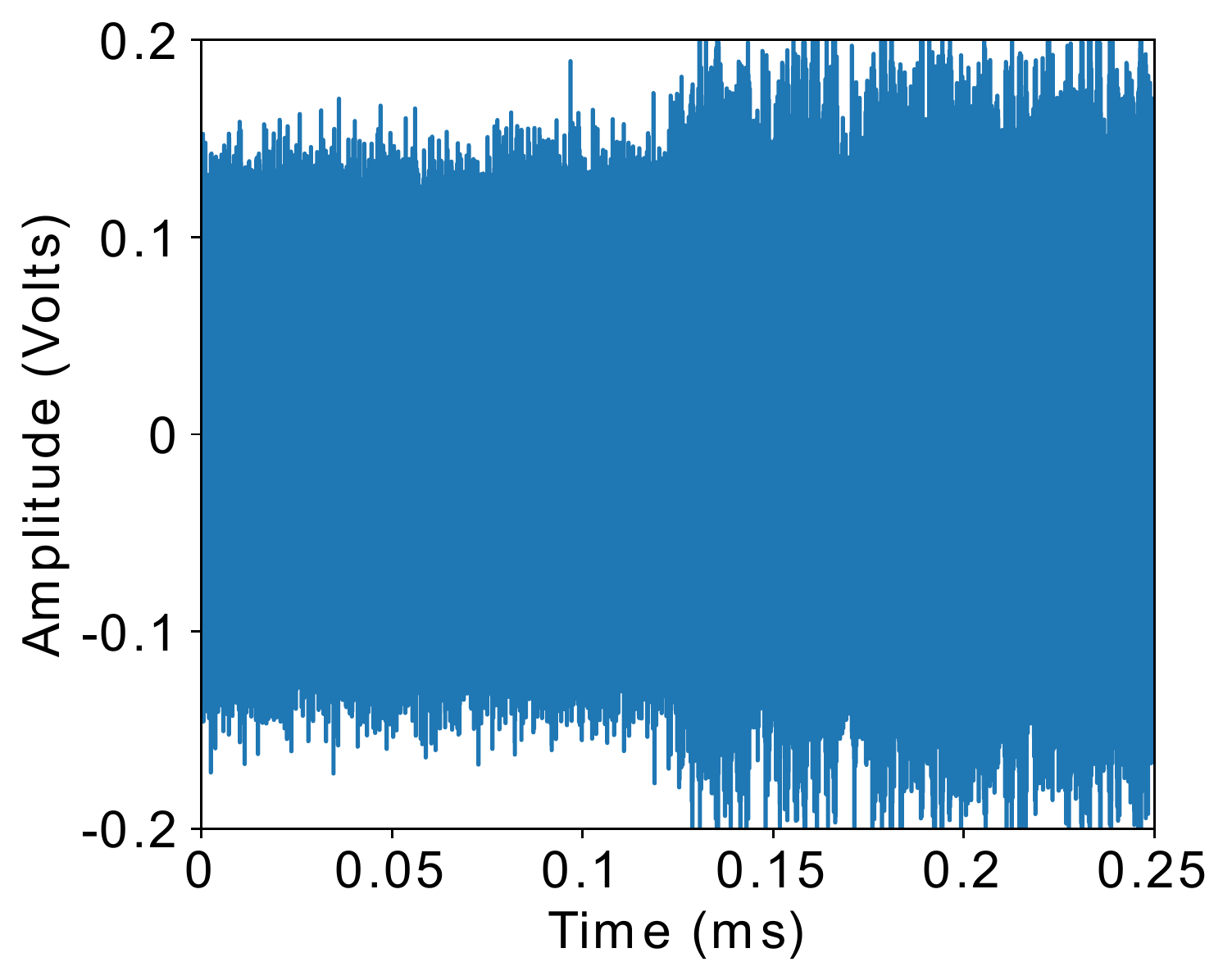}}
\hspace*{1cm} 
\vspace{-6mm}
\newline
\captionsetup[subfigure]{labelformat=parens}
\setcounter{subfigure}{0}
\subfloat[]{\includegraphics[width=0.3\linewidth]{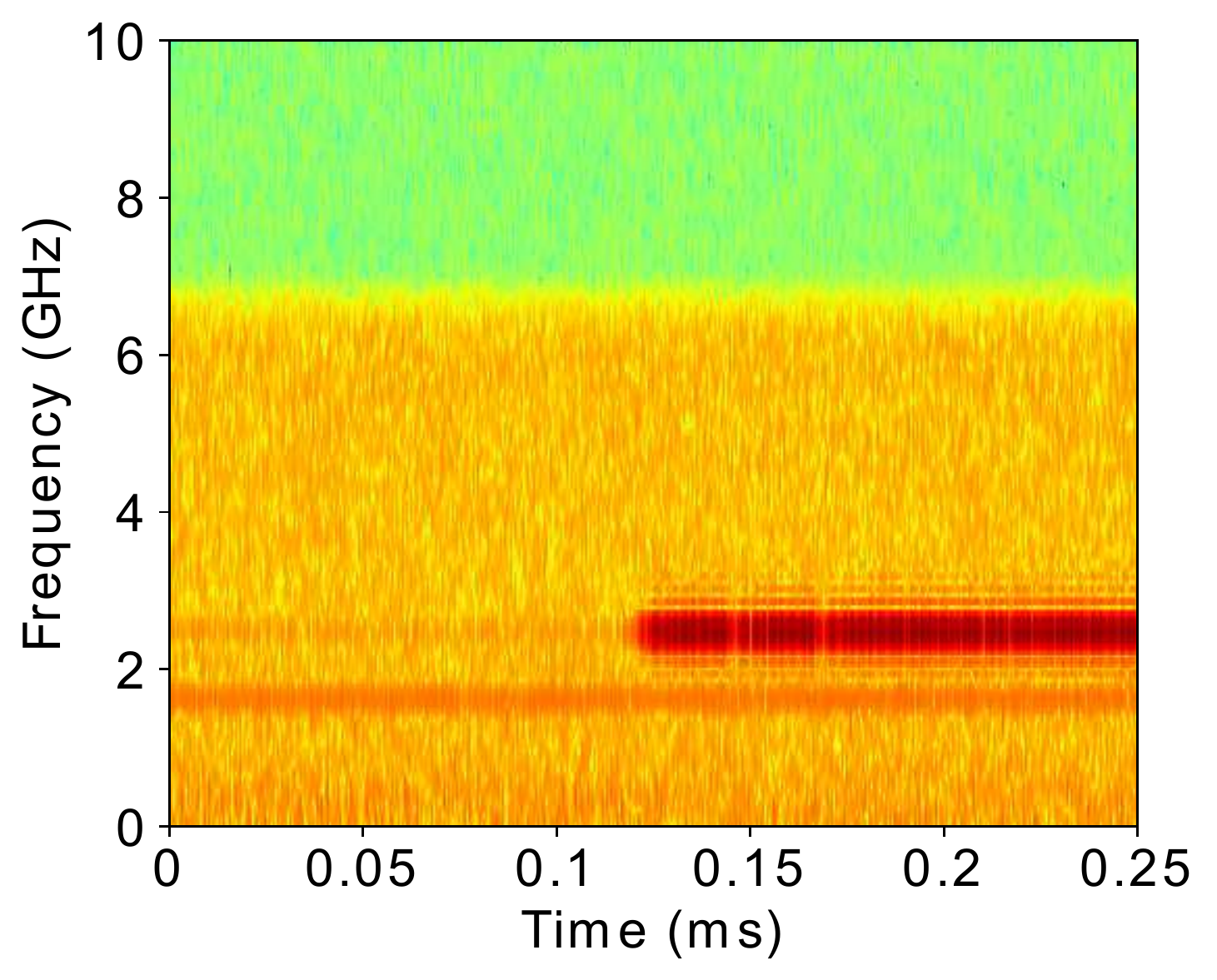}
\label{fig_noised_plots_a}}
\subfloat[]{\includegraphics[width=0.3\linewidth]{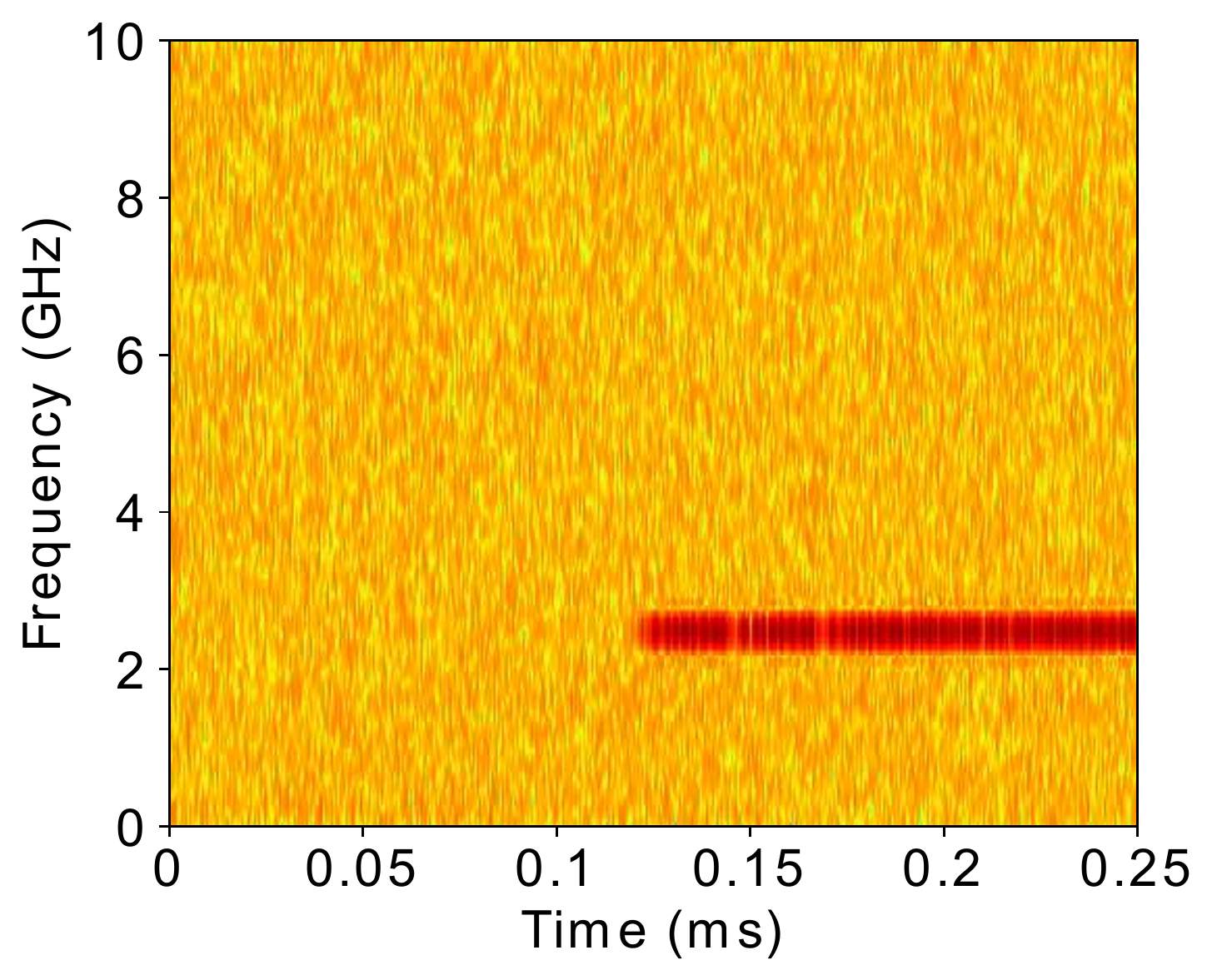}
\label{fig_noised_plots_b}}
\subfloat[]{\includegraphics[width=0.35\linewidth]{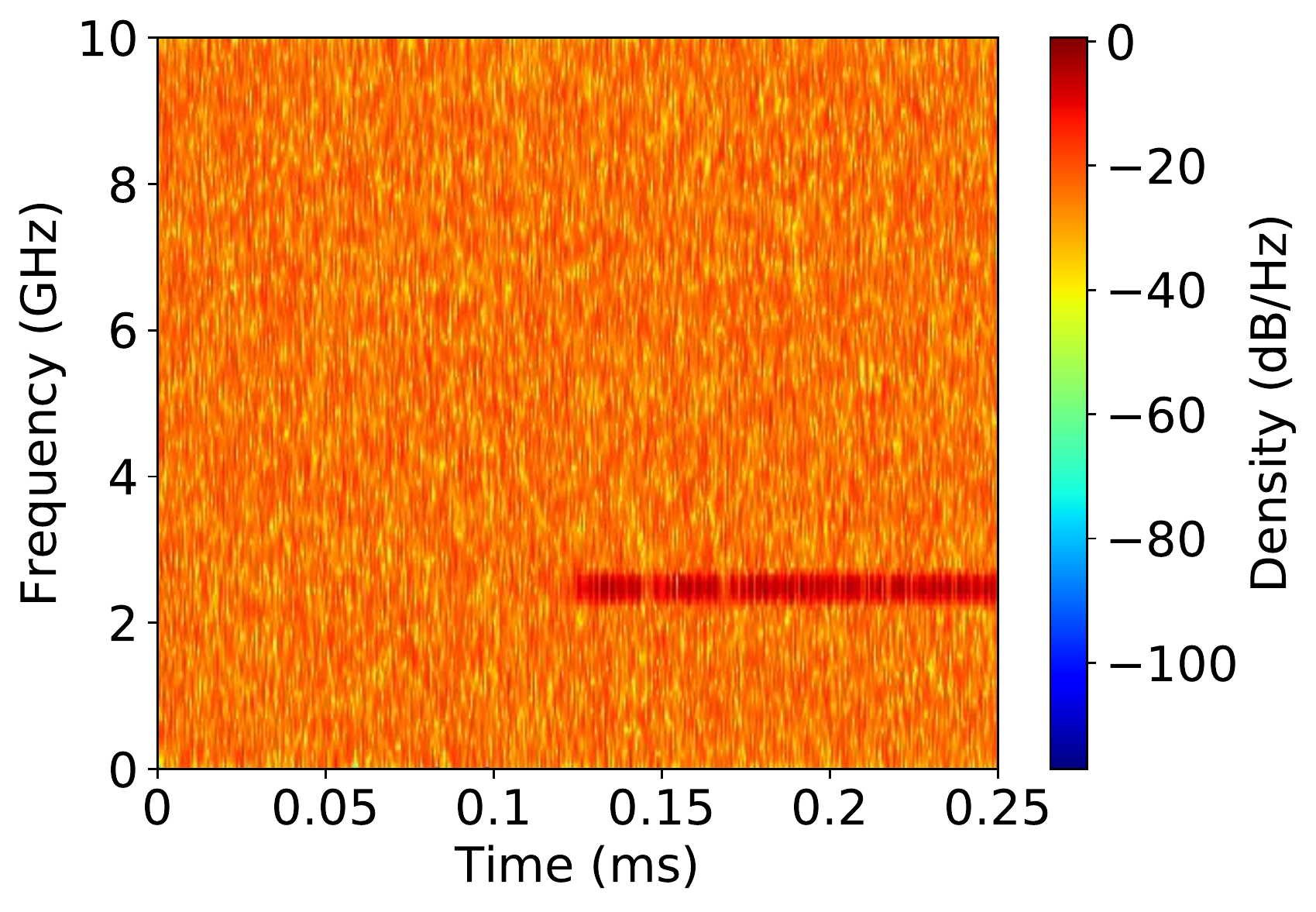}
\label{fig_noised_plots_c}}
\caption{Noisy controller signal of a DJI Inspire 1 Pro for different SNRs: a) 30~dB, b) 15~dB, and c) 0~dB. Base noise in lab environment is around 30~dB. As SNR decreases, distortion occurs for both image types. All spectrogram images have the same density color scale.} \label{Fig:noised_plots}
\vspace{-3mm}
\end{figure*}


\begin{table}
\renewcommand\arraystretch{1.3}
\caption{UAV controllers used in this work.}
\begin{center}
\begin{tabular}{cc}
\hline
UAV ID (\#) & Brand \& Model \\
\hline
\hline
1&Jeti Duplex DC-16\\
\hline
2&DJI Matrice 100\\
\hline
3&DJI Matrice 600\\
\hline
4&DJI Phantom 3\\
\hline
5&DJI Inspire 1 Pro\\
\hline
6&Spektrum DX5e\\
\hline
7&Spektrum DX6e\\
\hline
8&FlySky FS-T6\\
\hline
9&Futuba T8FG\\
\hline
10&Graupner MC-32\\
\hline
11&Hobby King HK-T6A\\
\hline
12&Spektrum JR X9303\\
\hline
13&DJI Phantom 4 Pro\\
\hline
14&Spektrum DX6i\\
\hline
15&Turnigy 9X\\
\hline
\end{tabular}\label{tab:controllers_list}
\end{center}
\vspace{-3mm}
\end{table}

Time-series RF signal of a controller is kept in a 1-D array. Time-series images are simply acquired by plotting these 1-D arrays. RF signals captured from different UAV controllers are illustrated in Fig.~\ref{Fig:sample_plots}. As it can be observed from the figure, RF signals exhibit different waveforms. Digital image processing literature bestows useful techniques to distinguish such signals using an envelope detector and template matching-based approaches~\cite{template_matching}. However, some controller signals may exhibit similar envelopes (e.g., RF signals in Fig.~\ref{Fig:sample_plots}(a) and Fig.~\ref{Fig:sample_plots}(b), or the signals in Fig.~\ref{Fig:sample_plots}(c) and Fig.~\ref{Fig:sample_plots}(d)), making it challenging to identify the controllers with these approaches. Besides, taking into account that signal envelopes get distorted at high noise levels, more advanced approaches are needed to achieve high classification accuracy. 

Spectrogram images are created calculating power spectral densities of the signals using Welch's average periodogram method, which is also called weighted overlapped segment averaging (WOSA) method~\cite{welch_newer}. In this method, time-domain signal $x[i]$ captured from a UAV is divided into successive blocks and averaged to estimate the power spectral density after forming the periodograms for each block, i.e.,
\begin{align}
    x_m[i] = w[i] x[i+mR]~,
\end{align}
where $i = 0,1,...,M-1$ is the sample index, $M$ is the window size, $m = 0,1,...,K-1$ denotes the window index, $K$ is the total number of blocks, $R$ is the window's hop size that tunes the amount of overlap between consecutive windows, and $w[i]$ is the window function. Then the periodogram of a block is calculated as
\begin{align}
    P_{x_m,M} (w_k) &= \frac{1}{M} |FFT_{N,k}(x_m)|^2\\ \nonumber
    &= \frac{1}{M}\left| \sum_{i=0}^{N-1} x_m[i]e^{-2\textrm{j}\pi ik/N} \right|.
\end{align}
Consequently, Welch estimate of power spectral density is calculated as follows
\begin{equation}\label{eq:Welch} 
    \hat{S}_x^{W}= \frac{1}{K} \sum_{m=0}^{K-1} P_{x_m,M}(w_k).
\end{equation}

In this paper Hanning window is used while calculating preiodograms. Then the calculated densities are mapped to a color scale to create spectrograms. We use a color map that spans the whole color space evenly, i.e., passes through all the colors in the visible range, which increases the accuracy of the proposed model significantly.
\subsection{Noising Procedure} \label{sec:noising_procedure}
Assuming fixed environmental noise, SNR level of an RF signal decreases as the source gets farther away from the receiving antenna. Since the drone or the controller position and hence their distance to the receiver antenna may vary in different scenarios, systems that can work under low SNR regimes are  required. In this study, we propose a method that can identify drones even at very low SNRs. Since the data~\cite{mpact_rffingerprints} is collected in a lab environment, the noise is stable and the same for all measurements. To train and test our models for noisy signals, we add white Gaussian noise to the raw data, and then generate the corresponding time-series and spectrogram images. While generating the noisy signals, we first use Higuchi's fractal dimension method~\cite{higuchi_physics_conf} to find the approximate position of the transient signal segment. We use this information to distinguish between the noise and the RF signal and calculate their actual power separately as 
\begin{equation}\label{eq:noising1} 
        P_{\textrm{noise}}\; [\textrm{dB}]  = 10\times \log_{10}\left( \frac{\sum_{i=0}^{T_b} |x[i]|^2}{T_b}\right),
\end{equation}
and 
\begin{equation}\label{eq:noising2} 
P_{\textrm{signal}}\; [\textrm{dB}] =10\times \log_{10}\left( \frac{\sum_{i=T_e}^{N} |x[i]|^2}{N-T_e}\right),
\end{equation}
where $T_b$ and $T_e$ are the indexes where the transient begins and ends, respectively.
\begin{figure*}[t]
\centering
\captionsetup[subfigure]{labelformat=parens}
\setcounter{subfigure}{0}
\subfloat[]{\includegraphics[width=0.33\linewidth]{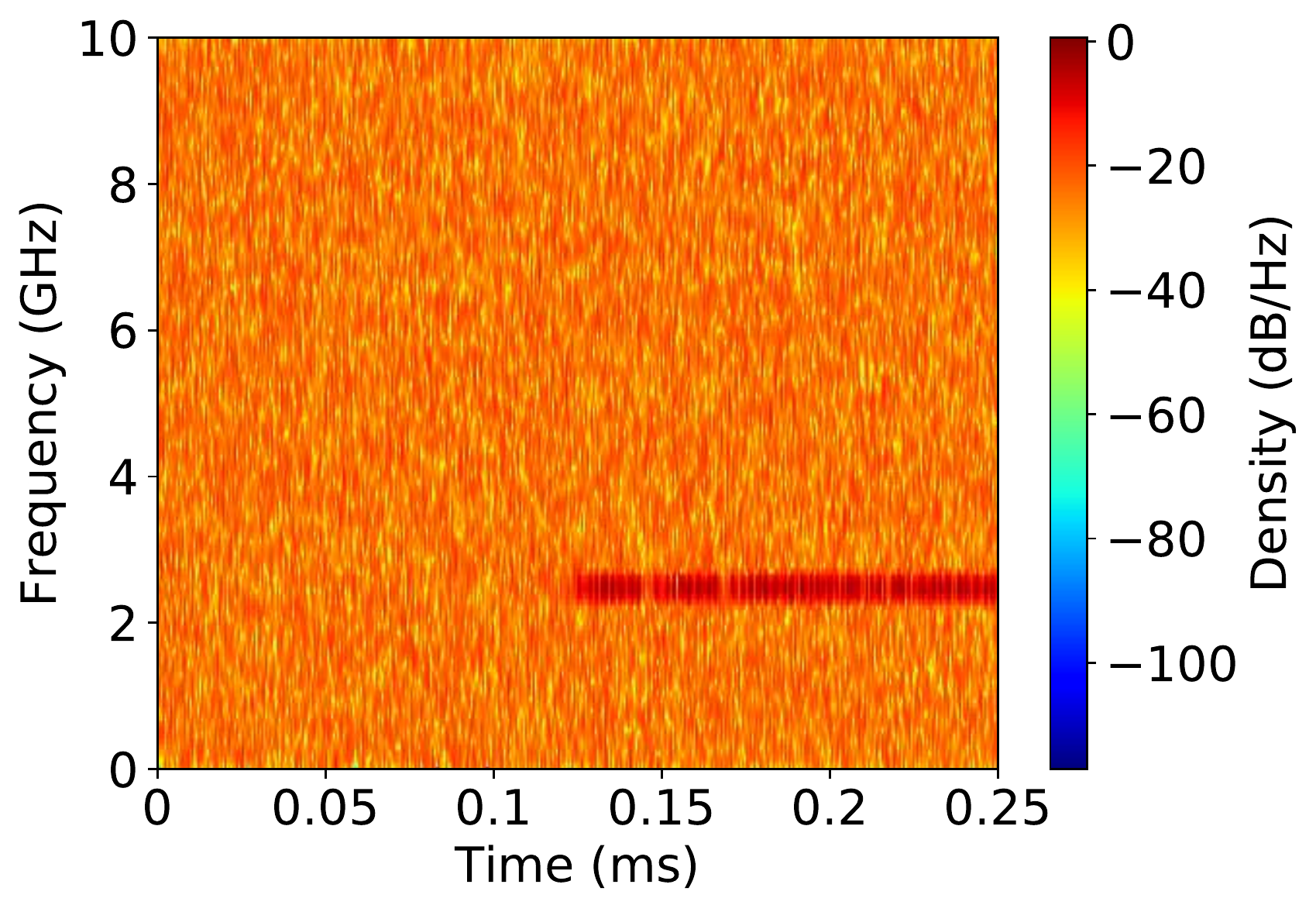}
\label{fig_denoised_plots_a}}
\subfloat[]{\includegraphics[width=0.33\linewidth]{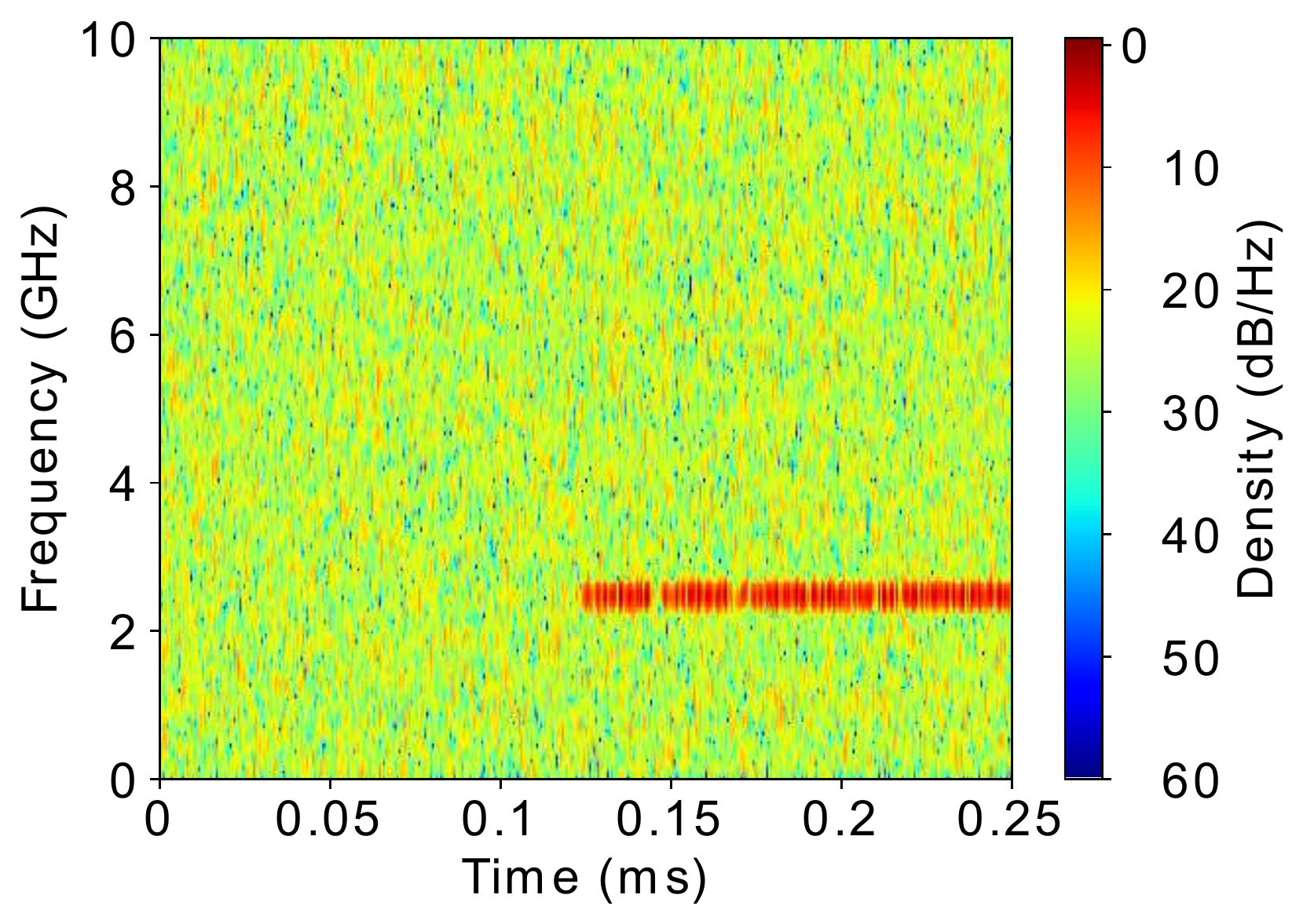}
\label{fig_denoised_plots_b}}
\subfloat[]{\includegraphics[width=0.33\linewidth]{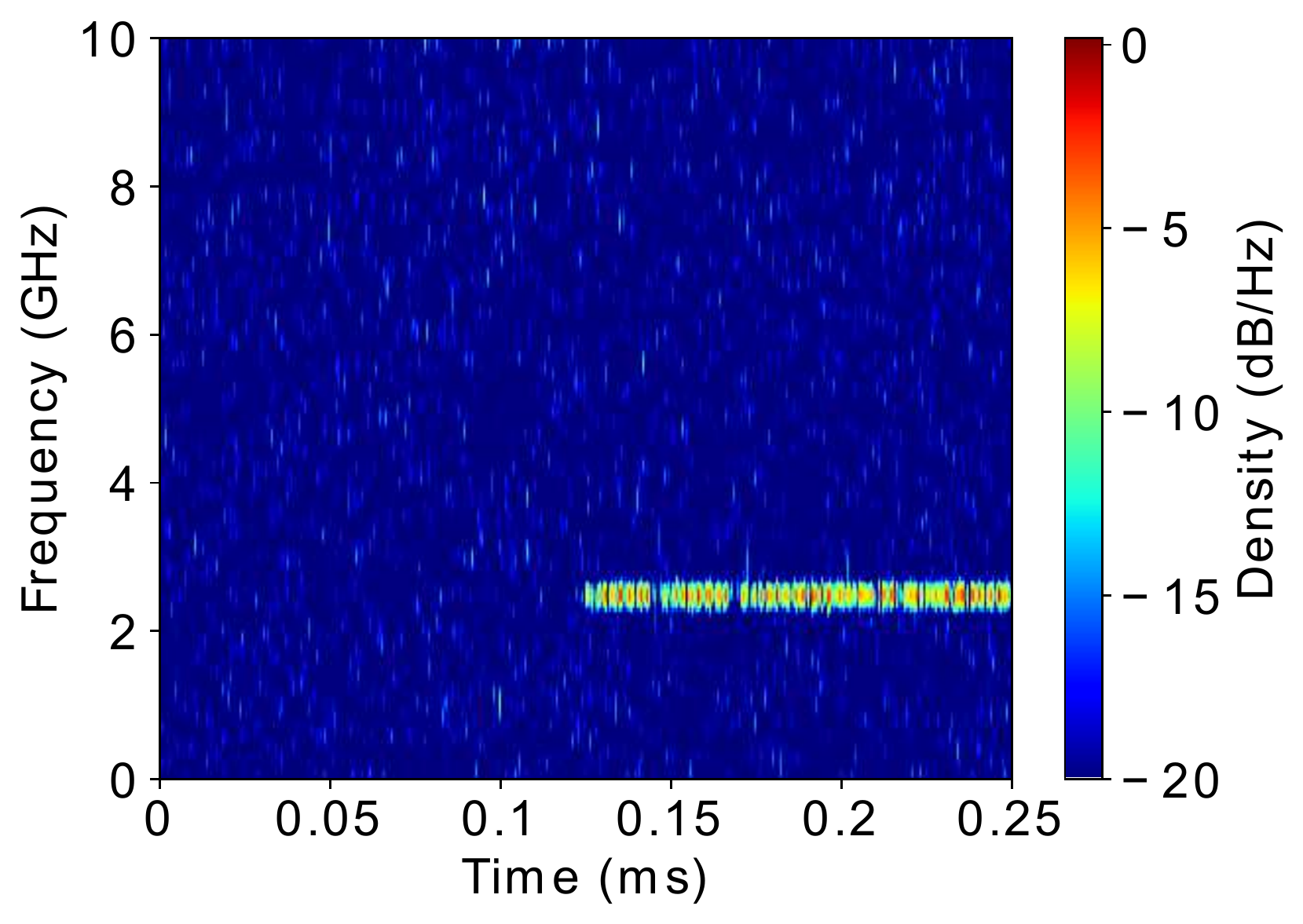}
\label{fig_denoised_plots_c}}
\caption{Denoising of a 0~dB SNR signal at different cut-off values: a) no truncation, b) -60~dB/Hz, and c) -20~dB/Hz. Truncation process sets the lower limit of the density color scales. This increases the level of representation of the high density signal components on the spectrogram image, consequently model accuracy is improved.} \label{Fig:truncation}
\vspace{-3mm}
\end{figure*}

Next, we calculate the SNR~($\Gamma$) of the received signal and the difference between the current SNR level and the desired SNR level as follows: 
\begin{equation}\label{eq:noising3} 
        \Gamma_{\textrm{signal}}\; [\textrm{dB}] = P_{\textrm{signal}}  - P_{\textrm{noise}},
\end{equation}
and 
\begin{equation}\label{eq:noising4} 
\Delta \Gamma\; [\textrm{dB}] = \Gamma_{\textrm{signal}}  - \Gamma_{\textrm{desired}}.
\end{equation}

Finally, appropriate amount of random noise $n[i]$ is added to the whole signal to set the signal to the desired SNR level as 
\begin{equation}\label{eq:noising} 
      s[i] = x[i] + n[i],
\end{equation}
where $n[i] = \Delta \Gamma \times \mathcal{N}(0,1), i = 0,1,..., N-1$, and $N$ is the total number of samples in $x[i]$. Note that $\Delta \Gamma$ that is used to generate the noise sequence is not in dB scale. 

A set of artificially noised time-series images and the corresponding spectrograms are given in Fig.~\ref{Fig:noised_plots}. Subject to the type of the controller, the original data has an SNR of about 30~dB. Increased noise causes distortion visible in both image types. However, time-series images are affected more. Spectrograms preserve signal characteristics better than time-series images as signal components can be better resolved in frequency domain.


\section{Image Preprocessing and CNN-Based UAV Classification}
CNN is a deep learning algorithm which has been proven to perform well in image recognition and classification tasks~\cite{geron2019hands}. CNNs have layers just as any other neural networks; however, different from other deep learning algorithms, convolution layers are used to apply various filters to an image to extract features no matter at which part of the image they reside. This nature of the algorithm makes CNN a perfect fit for 2D data (e.g., images), and also reduces the number of required weights in a neuron, thus yields lower computational complexity in comparison with conventional deep neural network architectures. In this work, spectrograms and time-series images of RF signals have been used as inputs to the CNN models.

Even though CNN is a very powerful approach for extracting features from images, preprocessing phase of the source data is crucial to increase the overall success of the classification and decrease the computational cost. 

\subsection{Conversion to Grayscale and Image Cropping}
As reviewed in Section~\ref{sec:Images}, spectrograms reflect the power spectral densities of the signals. Since color depth preserves distinctive information, these images should be kept in red/green/blue (RGB) format. However, time-series images are not represented in such a format, and therefore, to decrease the complexity, time-series signal images should be converted to grayscale if these images do not come in grayscale by default. \looseness=-1

Time-series and spectrogram images typically have axes, ticks and labels regardless of the software tool that is used to create them. We remove all those parts before beginning post-processing the images. Besides, captured images would include both the noise-only signal (when there is no transmission) and the transmitted signal (see the time-series signals in Fig.~\ref{Fig:noised_plots}). By using Higuchi's fractal dimension method as suggested in Section~\ref{sec:noising_procedure}, it is possible to remove out the noise-only part in both image types. In addition, one of the axes of the spectrograms will includes frequency domain information. In case the frequency range of interest is known, it is appropriate to crop the spectrograms further to lower the computational cost focusing on the desired frequency band only.

\subsection{Denoising the Spectrograms} \label{sec:denoising_the_spectrograms}

Denoising is an important step towards improving the accuracy of the spectrogram image-based classification. Denoising by truncation is only applied to spectrogram images. Power spectral densities should be calculated up to a certain frequency that is defined by the sampling rate for several instants in time domain that covers the whole signal. These spectral density values are mapped to the RGB color scale while creating spectrograms: the minimum and the maximum spectral densities are mapped to the coolest and warmest colors of a chosen color map, whereas the colors for the intermediate values are adjusted accordingly. In order to denoise the spectrogram, a cut-off density is picked as a threshold, and the spectrogram is truncated by setting the elements of the spectral density array that are smaller than this cut-off to the cut-off value itself. This process assures signal components with smaller densities to be cleared. Since most of the noise components have lower power densities than the drone controller signal itself for a wide range of SNRs, the truncation process is essentially a denoising procedure. Rest of the signal is mapped to the same color range set, which increases the level of representation of the details. As a result, non-noise (i.e., RF) signal components come forward that helps the CNN models learn better. 
\begin{figure*}[!t]
\centerline{\includegraphics[width=0.9\columnwidth]{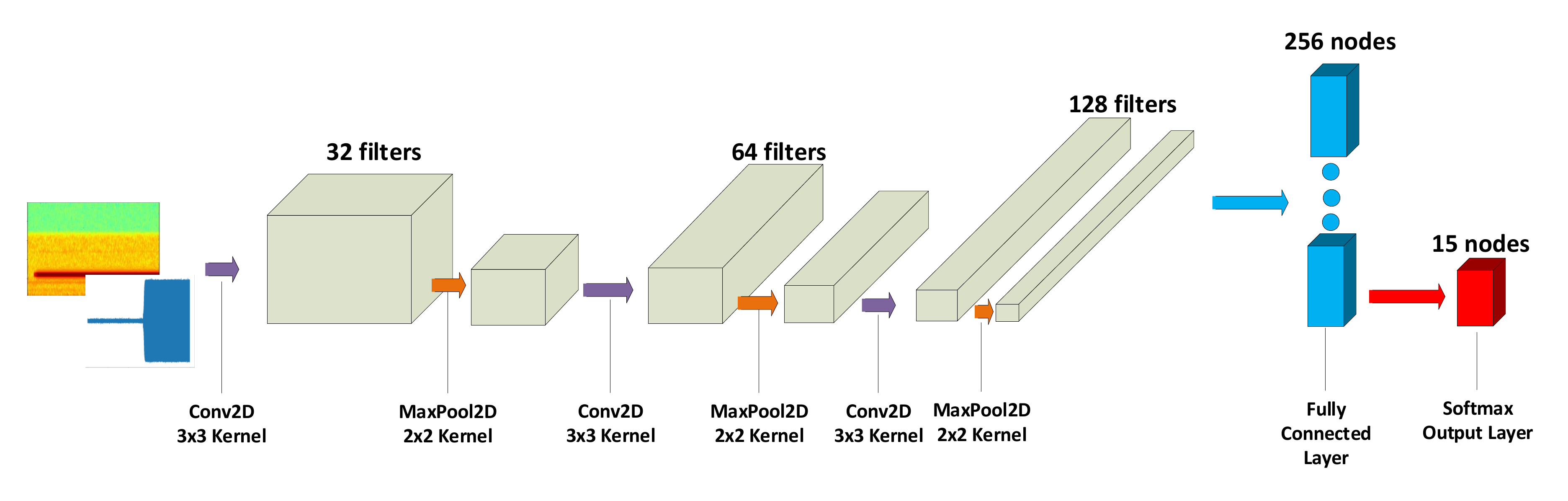}}
\caption{CNN architecture formed of three convolution and pooling layer pairs followed by a fully connected and a softmax output layer. Number of filters increase as data travels deep into the model to capture widening variety of features better. Softmax output layer gives a set of predictions resolving the maximum likelihood of the signals class reference. The one with the highest probability is predicted by the model to be the class of the signal. } \label{Fig:CNN_architecture}
\vspace{-3mm}
\end{figure*}
The procedure described above could be explained mathematically as follows:
\begin{align}
    S^{\prime}[f,i] &\xrightarrow[]{f_t}
        \begin{cases}
            \gamma_c, \quad& \text{if } S[f,i]\leq\gamma_c\\ 
            S[f,i], \quad& \textrm{else}
        \end{cases},\\
    S^{\prime}[f,i] &\xrightarrow[]{f_c} r_{f,i}, g_{f,i}, b_{f,i}, \nonumber
\end{align}
where $f_t$ is the truncation function, $S'[f,i]$ is the truncated signal subject to the cut-off value $\gamma_c$, $f_c$ is the color mapping function, and $r_{f,i}$, $g_{f,i}$, and $b_{f,i}$ are the color intensities in the corresponding channels.

There exists a critical trade-off that depends on the chosen cut-off threshold. For a given SNR level of the signal in hand, spectrograms should be truncated at an optimum level for that SNR. More specifically, in the case of under-denoising, excess noise causes overfitting, while in the case of over-denoising, useful information is wiped out together with the noise, which yields to underfitting. To illustrate this trade-off, we will consider Fig.~\ref{Fig:truncation} which shows the spectrograms of a DJI Inspire 1 Pro controller signal, that is artificially noised to 0~dB~SNR, at different truncation levels. In this figure, it is observed that as the threshold increases (i.e., from no truncation to $-$20~dB/Hz), lower limit of the density on the spectrograms changes. This lowest density is the lowest value in the domain set. As a result of truncation, high density components of the signals are represented better on the images.  

Another aspect of creating CNN models for different data sets at various SNR levels is the necessity of defining the SNR of a signal beforehand to invoke the appropriate model. 
To manage this, we propose to follow two approaches while working with the spectrograms. In the first approach, we create and optimize different CNN models for different SNR levels. The idea is that, assuming that the captured signal's SNR can be measured, the model having the closest SNR is called to perform classification. Even though, calculating the SNR of a received signal in real time is a tricky task, we believe, with the help of featured state-of-the-art measurement devices and newly developed algoritms~\cite{real_time_snr}, this would not be a problem. In the second approach, we define an optimum cut-off value (i.e., minimum average validation loss among all different cut-offs), merge all the images of different SNR levels truncated at this level to create a new comprehensive dataset, and then train a single model. The major advantage of this second approach is that it is no longer required to determine the SNR of the signal in advance.

\subsection{Training and Testing CNN Models}

In this work, CNNs are trained using Keras with Tensorflow at the backend. In the models created, we have three convolution layers (Conv2D) followed by pooling layers (MaxPool2D) and then a fully connected layer followed by the output layer. Convolution layers get deeper (i.e., the number of filters increase), and size of the images get smaller as the data travels deep into the model, in accordance with the general convention. The CNN models have been trained and tested with $3:1$ ratio for each UAV class. Optimum hyperparameters are determined after running vast amount of simulations. Results are presented in the next section. 

An illustration of the CNN architecture is shown in Fig.~\ref{Fig:CNN_architecture}. While training the models, categorical cross-entropy function is used as the loss function
\begin{align}
\label{eq:categorical_cross_entropy}
    \mathcal{L}(W)=-\frac{1}{N}\sum_{i=1}^{N}~&[y^{(i)}\log(\hat{y}^{(i)})+\\ \nonumber
    &(1-y^{(i)}) \log(1-\hat{y}^{(i)})], 
\end{align}
where $W$ represents the model parameters, and $y^{(i)}$ and $\hat{y}^{(i)}$ represent the true labels and predicted labels for the $i$-th image, respectively. This function gets smaller as the true and the predicted results get closer to each other. The aim of the model is to find the optimum set of model parameters to minimize this function, i.e.,
\begin{equation}
\label{eq:argmin}
   \Hat{W} = \argmin_W\mathcal{L}(W).
\end{equation}
Probability of the $i$-th test image, expressed as $\mathbf{x}^{(i)}$ in vector form, being a member of the $k$-th class is calculated using normalized exponential function as:
\begin{equation}
\label{eq:softmax}
    p_k(\mathbf{x}^{(i)}) = \frac{e^{\hat{v}_k^{(i)}}}{\sum_j e^{\hat{v}_j^{(i)}}}, 
\end{equation}
where $\mathbf{\hat{v}^{(i)}}$ is the $K \times 1$ vector output of the final model that uses optimized weights given in~(\ref{eq:argmin}), and $K$ is the number of classes. The class that has the maximum probability is chosen to be the prediction of the model for the $i$-th test image, $\hat{y}^{(i)}$, for the given image
\begin{equation}
\label{eq:estimate}
    \hat{y}^{(i)} = \argmax_k p_k(\mathbf{x}^{(i)}).
\end{equation}

The next section presents the experimental results that are acquired with the CNN models created for both the time-series and spectrogram images.

\section{Experimental Results}

During traning the CNN models, the original dataset in~\cite{mpact_rffingerprints}, where the SNR is about 30~dB for the whole set, was used. We extended this dataset by considering four additional SNR levels ranging from 0~dB to 20~dB for time-series signal-based classification, and seven different additional SNR levels ranging from $-$10~dB to 20~dB for spectrogram-based classification, with SNR increments of 5~dB in both cases. To train the models, we created 100 images for each class and for each unique SNR truncation threshold pairs following the noising procedure described in Section~\ref{sec:noising_procedure} and the denoising procedure (for spectrogram images only) described in Section~\ref{sec:denoising_the_spectrograms}. Throughout the study, we created more than 100 datasets, each having 1500 images (15 classes with 100 images each). A Hanning window function of size 128 with 16 overlap samples is used while creating the spectrograms.

Before feeding the CNN, we crop the images appropriately to get rid of the unnecessary parts of the images and reduce the file sizes, which helps speed up the converging of the CNN models. Resulting spectrogram and time-series signal images have the sizes of $(90\times385\times3)$ and $(779\times769\times1)$, respectively. In this work, we used brute force searching to optimize CNN model parameters. We utilized NC State University HPC (High Performance Computing) Facility to run parallel simulations for different sets of hyperparameters to find the optimum parameter set. 

In the rest of this section, we will first give considerations about the environmental interference issues and then present the classification results for the time-series images and spectrogram images. Subsequently, we will discuss the relation between classification accuracy and training set size and, finally, share the results for out-of-library classification performance of the proposed model.
\begin{table}
\renewcommand\arraystretch{1.15}
\caption{Optimum Set of Hyperparameters for Time-Series Images.}
 \begin{center}
\begin{tabular}{P{0.85cm}P{1.5cm}P{0.75cm}P{2cm}}
\hline
SNR (dB) & Optimizer & Batch size & Validation accuracy (\%) \\
\hline
\hline
30 &SGD&4&99.7\\
\hline
20&Adagrad&4&96.5\\
\hline
10&Nadam&16&81.6\\
\hline
5&Nadam&1&65.3\\
\hline
0&Adagrad&1&50.1\\

\cline{1-4}

\end{tabular}\label{tab:R_results}
\end{center}
\vspace{-3mm}
\end{table}

\begin{table}
\renewcommand\arraystretch{1.15}
\caption{Optimum Set of Hyperparameters for Spectrogram Images.}
\begin{center}
\begin{tabular}{P{0.85cm}P{1.6cm}P{1.5cm}P{0.75cm}P{2cm}}
\hline
SNR (dB) &Cut-off level (dB/Hz)& Optimizer & Batch size & Validation accuracy (\%) \\
\hline
\hline
30 &$-$100&Adamax&8&99.7\\
\hline
20&$-$90&Nadam&2&99.7\\
\hline
15&$-$10&Adam&32&100.0\\
\hline
10&$-$10&Nadam&2&100.0\\
\hline
5&$-$10&Adam&4&99.7\\
\hline
0&$-$10&Nadam&8&99.5\\
\hline
$-$5&$-$20&RMSProp&8&99.5\\
\hline
$-$10&$-$15&Nadam&16&92.0\\
\hline
Merged&$-$10&SGD&1&98.8\\
\hline
Merged\mbox{*}&$-$10&SGD&1&96.9\\
\cline{1-5}
\multicolumn{5}{l}{\renewcommand\arraystretch{0.7}\mbox{*}Refers to the set of images created by assuming SNR levels}\\
\multicolumn{5}{l}{\renewcommand\arraystretch{0.7}different than the ones used to train the merged model.}\\
\end{tabular}\label{tab:S_results}
\end{center}
\vspace{-3mm}
\end{table}

\subsection{Comments on Environmental Interference}
All the signals used in this study are recorded for a wide range of frequencies, i.e., 0$-$10~GHz, as illustrated by the spectrogram in Fig.~\ref{Fig:noised_plots}(a). The first observation that can be made in there is that the frequency utilization significantly decreases above roughly 7~GHz, which is because there is no wireless transmission for that frequency range nearby the locations where we conducted the measurements. One can also notice the high color intensity at GSM band around 1800~MHZ. Since all of the drone controllers considered in this study transmit in 2.4~GHz ISM band, notable densities in other bands on spectrograms have no effect on the model accuracy. However, 2.4~GHz band is also used heavily by Wi-Fi and Bluetooth transmitters. In case Wi-Fi and/or Bluetooth signals are received, our proposed model applies a multistage detection system described in~\cite{mpact_rffingerprints} to detect those type of signals and filter them out. 

Raw data used in this work have been gathered in an indoor environment where Wi-Fi and Bluetooth signals could exist. A 24~dBi gain directional antenna has been used to capture the signals. It is known that IEEE~802.11 standards family routers implement carrier-sense multiple access (CSMA) techniques, which may help reducing the probability of interference with Wi-Fi transmitters when the drone controllers are close to the receiving antenna. Besides, low-power Bluetooth transmitters will not possess high risk of severe distortion on the received signal. Moreover, our classifier makes a decision each time after processing a signal frame of 250~$\mu$s.
Short duration of signals allow our system to catch drone controller signals even in the existence of other packet-based communication technologies as they do not transmit packets continuously. While capturing a drone data-only signal frame may introduce time delays in identifying the drone, this delay will be on the order of milliseconds. Therefore, we can safely conclude that labeling the training set as if there are no WiFi and Bluetooth signals complies with real-world scenarios.

\subsection{UAV Classification Using Time-Series Images}\label{sec:results_timeseries}
We optimize five different CNNs for time-series images by brute force searching approach. We ran simulations for each dataset using all combinations of seven different optimizers, seven different batch sizes, and five different activation functions, which add up to 245 distinct simulations. The parameter set that gives the highest accuracy is chosen. Optimized parameters for these models are given in Table~\ref{tab:R_results} for reproducibility. We observe that CNNs gather distinctive features from both the transient (i.e., the signal segment where the noise-only region ends and the RF signal begins) and the envelope of the RF signal. As the signal swamps into noise as SNR decreases, first the transient information disappears whereas the information carried in the signal envelope survives a little longer. When the SNR is further decreased, envelope information also disappears. Thus, the validation accuracy drops from 99.7\% to 50.1\% as the SNR goes down from around 30~dB to 0~dB. Even though different optimizers could give the maximum accuracy for different SNR levels, all optimum models use rectified linear unit~(ReLu) as the activation function. 

Both in-band and out-of-band noise cause distortion in time-series images of RF signals, and so the models trained on time-series images suffer from noise more than the models that use the spectrogram images. Besides, while using the time-series images, trained models extract features from the amplitude of the signals itself. However, amplitude of a received signal depends on the distance between the receiver and transmitter antenna. This is an obvious problem when the only distinctive difference between the time-series signal images of any two controllers is the difference in their amplitude (e.g., see the RF signals in in Fig.~\ref{Fig:sample_plots}(c) and Fig.~\ref{Fig:sample_plots}(d)). Models trained on spectrograms show better performances. Next subsection is dedicated to results of models that employ spectrogram images.


\begin{figure}[t]
\centerline{\includegraphics[trim=0cm 0cm 0cm 1.7cm, clip,width=.75\columnwidth]{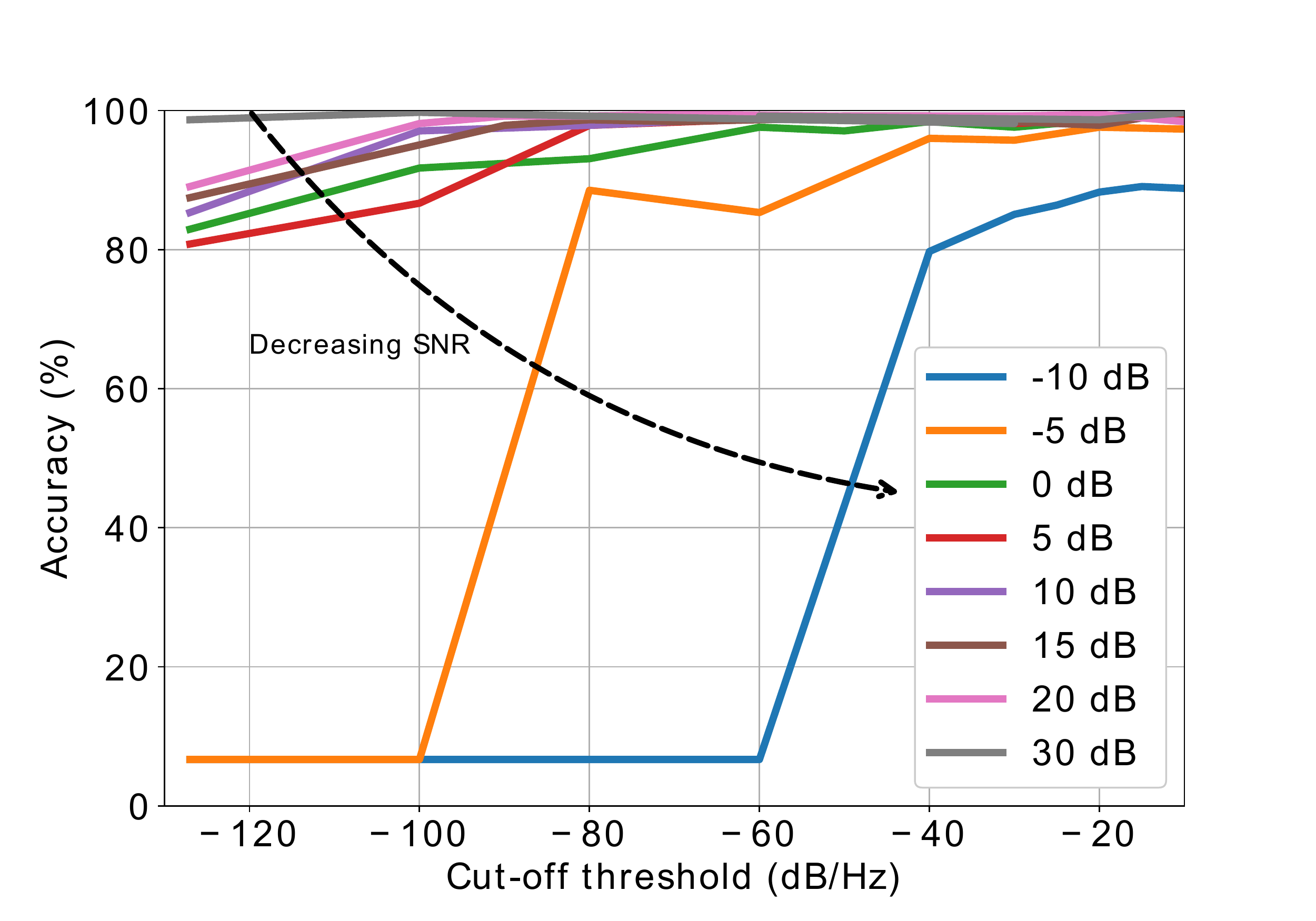}}
\caption{Spectrogram model classification accuracy versus the cut-off threshold for different SNR levels. Denoising the spectrograms by truncating the spectral densities, subject to a threshold, increases the model accuracy in general. Models trained with high-SNR data give reasonable accuracy even without denoising. Low-SNR models need to be denoised a priori.} \label{Fig:cutoff_vs_accuracy}
\vspace{-3mm}
\end{figure}

\subsection{UAV Classification Using Spectrogram Images} \label{sec:results_spectrograms}
Two approaches are adopted while creating models on spectrogram images. In the first approach, we assumed that the SNR level of a received signal can be measured prior to classification and created different models for different SNR levels. In the second approach, we used a merged dataset that includes spectrogram images of different SNRs to create a model that can be used to classify any received signal without any prior information about its SNR. Details of these approaches are given in the following subsections. 

\subsubsection{Models with Single SNR Training Sets}
We have created eight models for eight different SNR levels that are truncated at their own optimum levels. To use this approach, the SNR of the received signal should be calculated first, and then the model that has the closest SNR should be called to perform the classification. 
We observed that all of the models give their highest accuracy with the ReLu activation function. The sensitivity of the validation accuracy to a single output was found to be 0.27\%~sample$^{-1}$. Optimized parameters of the models using spectrogram images are given in Table~\ref{tab:S_results}. It is seen that the lowest accuracy belongs to the SNR level $-$10~dB among the individual sets. Performances of all the other models can be considered almost perfect. It is also observed from Table~\ref{tab:S_results} that the optimum cut-off levels are different for different SNR levels. 

Classification accuracy at different truncation thresholds for different SNR levels are given in Fig.~\ref{Fig:cutoff_vs_accuracy}. By considering this figure and Table~\ref{tab:S_results} together, one can conclude that, in general, the classification accuracy tends to increase with the increasing level of truncation. For high SNRs (i.e., 20~dB and 30~dB), spectral densities of the signals are much higher than the noise; therefore, truncating the images at different levels does not wipe out much information. As a result, the accuracy curve navigates flatter, and the necessary cut-off threshold is low ($-$100~dB/Hz and $-$90~dB/Hz). At medium SNRs (i.e., 0$-$15~dB), high level of truncation is required to preserve as much information as possible (all $-$10~dB/Hz). On the other hand, at the lowest end of SNRs (i.e., $-$5~dB and $-$10~dB), without truncating the images, no learning occurs at all. For these lowest two SNRs, distinctive information in the spectrograms is swamped into noise so with no truncation, the accuracy is found to be only 6.66\%. As the cut-off threshold increases, first a reasonable accuracy is acquired for $-$5~dB SNR dataset at $-$80~dB/Hz threshold level. This amount of filtering is still not sufficient for $-$10~dB SNR, which only begins to learn at a comparably higher threshold of $-$40~dB/Hz. Moreover, $-$10~dB/Hz threshold level gives lower accuracy than the models trained at medium SNRs (i.e., 0$-$15~dB) using the same threshold. This is because over-denoising chops the meaningful information together with the noise, and consequently, the optimum cut-off level is slightly lower than $-$10~dB/Hz (i.e., $-$20~dB/Hz for $-$5~dB SNR and $-$15~dB/Hz for $-$10~dB SNR). If cut-off threshold is too high, this wipes out all the information, making all spectrograms look alike and consequently, there will be no learning.

The advantage of using spectral domain information could be seen from the results of 0~dB SNR model where the classification accuracy for time-series images is only 50.1\% (Table~\ref{tab:R_results}), whereas it is 82.9\% (Fig.~\ref{Fig:cutoff_vs_accuracy}) for the spectrogram model at the same SNR level.



\begin{figure}[!t]
\centerline{\includegraphics[width=.75\columnwidth]{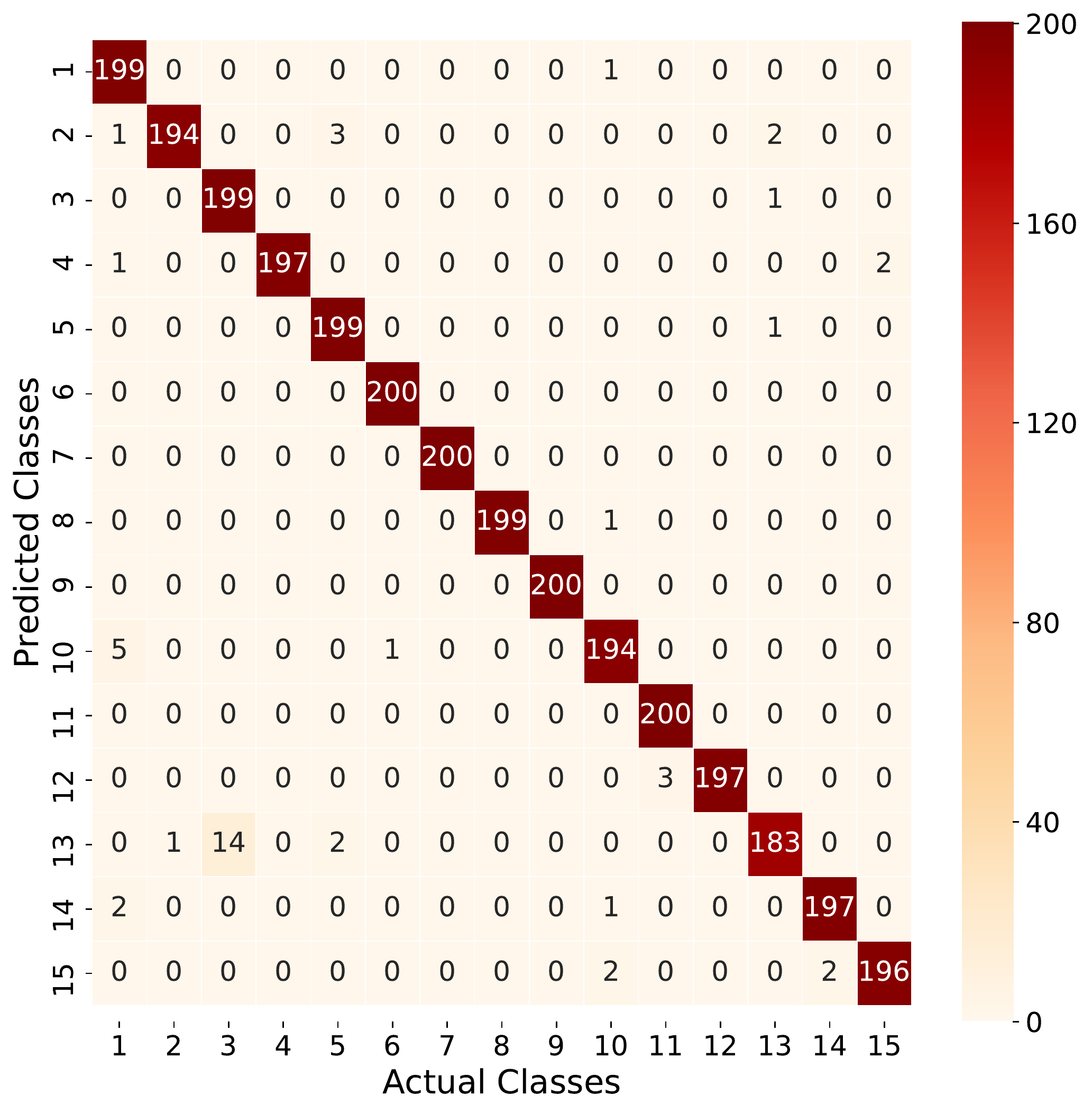}}
\caption{Confusion matrix for the merged model. Diagonal elements represent true positives, and off-diagonal elements represent the confusion between the classess.} \label{Fig:confusion}
\vspace{-3mm}
\end{figure}

\begin{figure}[!t]
\centerline{\includegraphics[trim=0.0cm 0cm 2cm 1.8cm, clip,width=.75\columnwidth]{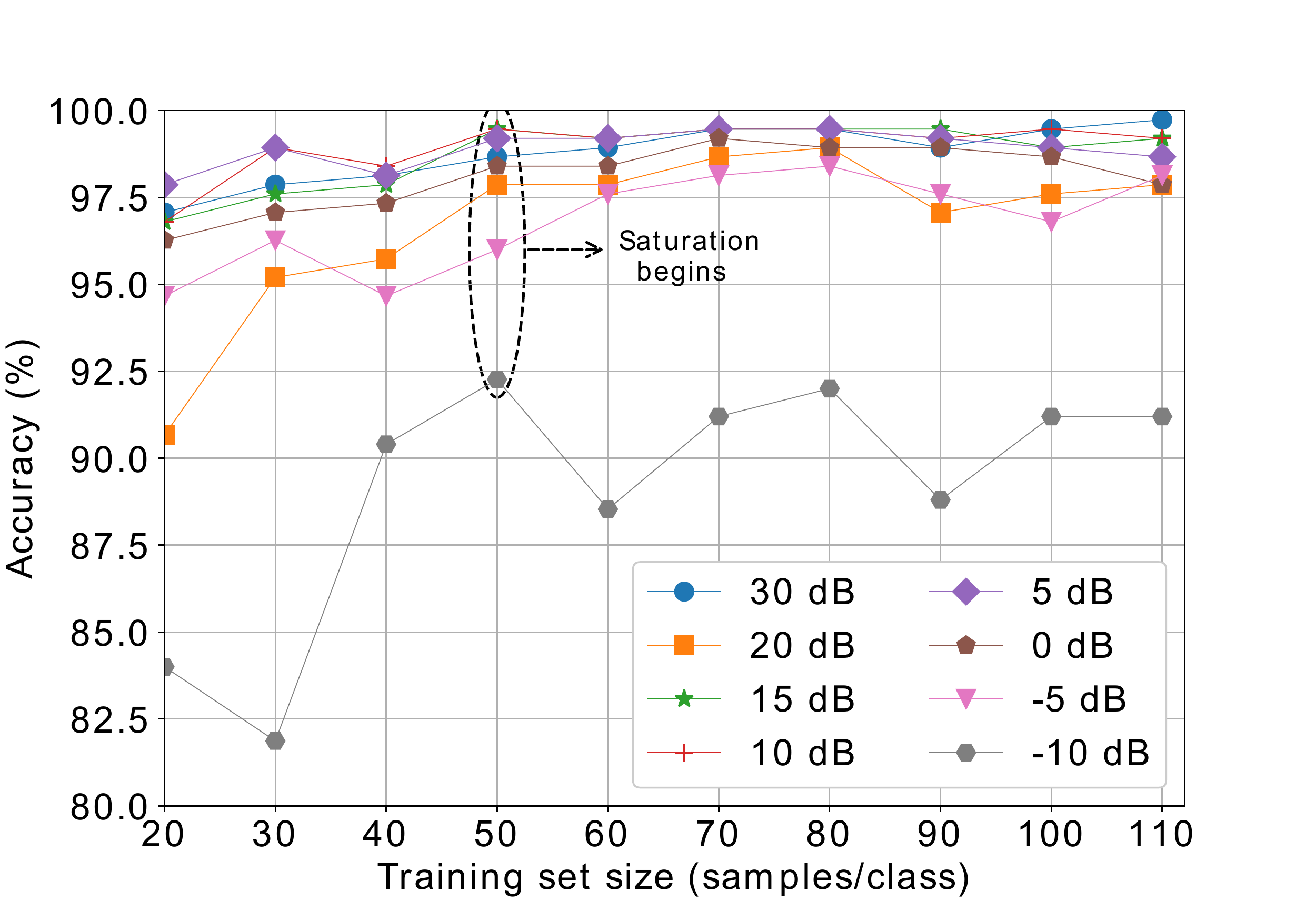}}
\caption{Classification accuracy as a function of training set size. Models give reasonable accuracies even with very low training data sizes. Saturation begins after 50 samples/class. We used 75 samples/class throughout this work.} \label{Fig:train_size_vs_acc}
\vspace{-3mm}
\end{figure}

\subsubsection{Model with a Merged Training Set}
Even though the models trained with different single-SNR data sets give satisfactory results, this approach comes with a practical difficulty. We can use these models only if we can measure the SNR of a received signal prior to classification. In order to get rid of this requirement and also to save time, we merged the training sets of different SNR levels to create a more generalized model. The truncation cut-off that gives the smallest average loss was found to be $-10$~dB/Hz, therefore we merged all the images for eight SNR levels denoised at this cut-off threshold. Training and test sets are eight times greater than those of the single SNR sets. A classification accuracy of 98.8\% (across all SNR levels) is achieved when using this model.

Confusion matrix of the merged model is given in Fig.~\ref{Fig:confusion}. The major deficiency of the model is observed at $(13,3)$, where 14 out of 200 test data which belongs to class 3 is predicted as class 13. These two controllers belong to the same company and both their time-series plots and spectrograms show high virtual resemblance.

We also tested the merged model with images at intermediate SNR levels ranging from $-$12~dB to 22~dB with increments of 5~dB. We used 30 images for each class at each SNR, which add up to 3600 images, all previously unseen to the classifier, to test the model. Our model gives 96.9\% accuracy, as shown in Table~\ref{tab:S_results}. In the case when we exclude the test data at $-$12~dB, accuracy of the model increases up to 99.3\%, which indicates that almost all the misclassification is associated with this particular SNR level. 

\subsection{Classification Accuracy vs. Training Set Size}
There are popular CNN models in the literature that can be implemented to a wide variety of image classification problems via transfer learning, e.g., VGG16 or InceptionV3. These models have abundant hidden layers and have been trained over enormous data sets. Other than these models, it is more customary to come across CNN models that are deeper and trained on larger data sets in the literature. If the problem in hand is to accurately classify images of miscellaneous objects, e.g., humans, animals or cars, then a deep model with a very large number of training set should be required. This is because these images have more diversity in terms of position, angle, ambiance, lightning, etc. However in our model, the set of images that we classify are generated by the well-defined methods that use the outputs of quite robust electronic circuitry. Thus, proposed models reach to very high accuracy with as low as 100 training samples per class. 

To better explain the sufficiency of low number of samples for this particular problem, we examined the dependence of accuracy to the training data set size. Fig.~\ref{Fig:train_size_vs_acc} shows the accuracy of the classifier with respect to sample size per class for different SNR levels. In these simulations, same models that are optimized for 100 samples per class are used for all cases. Note that, in this figure, x-axis denotes the size of the training sets only. We did not shrink the validation set while we tune the training set sizes. All models have been validated with a test set of 25 samples per class unseen by the models before. Here it is seen that for small training set sizes, classification accuracy decreases as expected. After roughly 50 samples/class, the accuracy reaches saturation and begins to fluctuate. On the other hand, we see that the created models give reasonable accuracy even for a training set size of as low as 20 samples per class, which is because these samples are created by devices that have high level of consistency.

For the practicality of the proposed system, the RF signal database should be updated as new products are introduced to the market. This also requires retraining of the CNN models, and hence fast training algorithms are needed. However, as explained above, the proposed system only requires a limited amount of training data, which in turn makes it a promising solution.


\begin{figure}[!t]
\centering
\centerline{\includegraphics[trim=0.3cm 0cm 1.3cm 1.6cm, clip,width=.75\columnwidth]{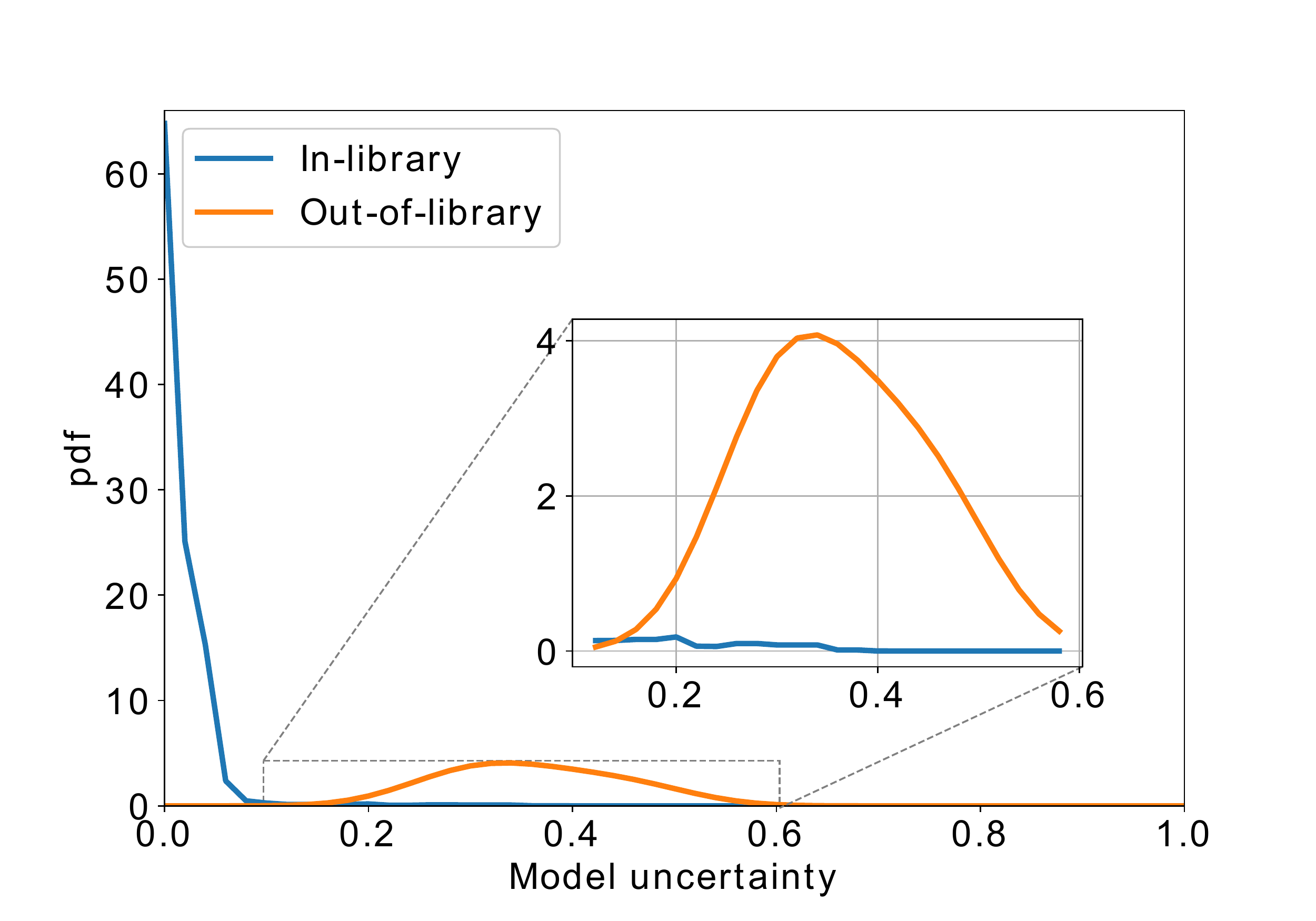}}
\caption{Pdfs of the model uncertainty for in-library and out-of-library UAV classes. The model predicts in-library signals with high certainty. The prediction uncertainty increases when the model encounters an out-of-library controller.} \label{Fig:out-of-library}
\vspace{-3mm}
\end{figure}

\subsection{Out-of-Library UAV Controller Signals}
Finally, we investigate the behavior of the proposed algorithm when the receiver captures an out-of-library UAV controller signal. To do that, we tested our optimized CNN-based classifier for 40 signals from Hubsan H501S X4 drone and compared the estimated probability distribution functions~(pdfs) of the prediction uncertainty with those of the in-library test signals in Fig.~\ref{Fig:out-of-library}. The output layer of the trained model gives a set of predictions for an incoming signal, where each element of the set corresponds to the estimated probability of that signal belonging to a particular class. Final decision on the class of the test signal is made based on the maximum probability, $p_{\max}$, in this set. We define the model uncertainty in Fig.~\ref{Fig:out-of-library} as $(1-p_{\max})$. 

We observe that the two classes (i.e., in-library and out-of-library UAVs) are well-separated in terms of the model uncertainty associated with each, and out-of-library UAV signals can be easily identified by a simple thresholding mechanism. The threshold can be placed based on the system requirements, i.e., the desired classification performance and false alarm rate. We recognize that a complete consideration of out-of-library classification requires adding out-of-library data in training set or adaptation of open set recognition by introducing an OpenMax layer, which estimates the probability of an input being from an unknown class~\cite{open_max}. On the other hand, our proposed model gives very low model uncertainty for in-library signals, and this therefore still provides a reasonable 
solution described above to detect the out-of-library drones in practice.
%

\section{Conclusion}
In this study, we proposed a system that uses drone controller RF signals to classify drones of different makes and models for a wide variety of SNRs. We used CNN classifiers with two different sources for training the models: time-series images, and spectrogram images.  
We showed that the CNN model using the spectrogram images is more resilient to noise 
when compared with the time-series images based model.
The proposed method that uses a merged training set of RF signals at different SNR levels along with the proposed denoising mechanism was shown to be effective for UAV classification even at SNRs not directly considered by the trained model. We also explored classification performance against training set size and showed that reasonable classification accuracy can still be obtained with limited training data. Consequently, adding new classes to the model (e.g., to include data from newly released drones) does not entail a high computation cost. 
Finally, we examined the model behaviour with in-library and out-of-library drone signals and concluded that the proposed model shows a good performance identifying drones from an unknown class. Our future work includes testing of the proposed CNN-based UAV classification technique scale, such as using the AERPAW experimental platform.
\section*{Acknowledgment}
This work has been supported in part by NASA under the Federal Award ID number NNX17AJ94A. The authors would like to thank Martins Ezuma at NC State for providing the drone controller RF dataset used in this study. 

\bibliographystyle{IEEEtran}
\bibliography{IEEEabrv,papers}
\vspace{-0.9cm}

\vfill
\end{document}